\documentclass[fleqn,usenatbib,useAMS]{mnras}

\usepackage{graphicx}	
\usepackage{amsmath}	
\usepackage{amssymb}	
\usepackage{multicol}        
\usepackage{bm}		
\usepackage{pdflscape}	
\usepackage{xcolor}
\usepackage{listings}

\DeclareRobustCommand{\Sec}[1]{Sec.~\ref{#1}}

\DeclareRobustCommand{\App}[1]{App.~\ref{#1}}

\DeclareRobustCommand{\Fig}[1]{Fig.~\ref{#1}}

\definecolor{SeaGreen}{rgb}{0.13,0.55,0.33}

\def\beq{\begin{equation}}
\def\eeq{\end{equation}}
\newcommand{\bea}{\begin{eqnarray}\begin{aligned}}
\newcommand{\eea}{\end{aligned}\end{eqnarray}}

\newcommand{\Gaia}{\textit{Gaia}~}

\usepackage[T1]{fontenc}
\usepackage{ae,aecompl}

\usepackage{newtxtext,newtxmath}

\title[Via Machinae]{Via Machinae: Searching for Stellar Streams using Unsupervised Machine Learning}

\author[Shih et al.]{
David Shih,$^{1}$\thanks{E-mail: shih@physics.rutgers.edu}
Matthew R.~Buckley,$^{1}$
Lina Necib,$^{2,3,4}$
and John Tamanas$^{5}$
\\
$^{1}$NHETC, Dept. of Physics and Astronomy, Rutgers, Piscataway, NJ 08854, USA\\
$^{2}$Walter Burke Institute for Theoretical Physics,
California Institute of Technology, Pasadena, CA 91125, USA\\
$^{3}$Center for Cosmology, Department of Physics and Astronomy,
University of California, Irvine, CA 92697, USA\\
$^{4}$Observatories of the Carnegie Institution for Science, 813 Santa Barbara St., Pasadena, CA 91101, USA\\
$^{5}$Department of Physics, University of California Santa Cruz, 1156 High Street, Santa Cruz, California 95064, USA
}

\pubyear{2021}

\begin{document}
\label{firstpage}
\pagerange{\pageref{firstpage}--\pageref{lastpage}}
\maketitle

\begin{abstract}
We develop a new machine learning algorithm, \textsc{Via Machinae}, to identify cold stellar streams in data from the {\it Gaia} telescope.
\textsc{Via Machinae} is based on ANODE, a general method that uses conditional density estimation and sideband interpolation to detect local overdensities in the data in a model agnostic way. By applying ANODE to the positions, proper motions, and photometry of stars observed by {\it Gaia}, \textsc{Via Machinae} obtains a collection of those stars deemed most likely to belong to a stellar stream. We further apply an automated line-finding method based on the Hough transform to search for line-like features in patches of the sky. 
In this paper, we describe the \textsc{Via Machinae} algorithm in detail
and demonstrate our approach on the prominent stream GD-1. 
Though some parts of the algorithm are tuned to increase sensitivity to cold streams, the \textsc{Via Machinae} technique itself does not rely on astrophysical assumptions, such as the potential of the Milky Way or stellar isochrones.
This flexibility suggests that it may have further applications in identifying other anomalous structures within the {\it Gaia} dataset, for example debris flow and globular clusters. 
\end{abstract}





\defcitealias{2018ApJ...863L..20P}{PWB18}
\defcitealias{Nachman:2020lpy}{NS20}
\defcitealias{full_sky}{Paper II}

\section{Introduction}

Stellar streams, the tidally-stripped remnants of dwarf galaxies and globular clusters, provide a unique window into the properties of the Milky Way and its formation history. Streams trace the
historical record of the mergers that built the Milky Way \citep{1998ApJ...495..297J,1999MNRAS.307..495H,2006ApJ...642L.137B,2018Natur.563...85H,2018MNRAS.478..611B,2021arXiv210409523M}. 
Their orbits allow measurements of the underlying gravitational potential of the Milky Way \citep{1999ApJ...512L.109J,2001ApJ...551..294I,2010ApJ...712..260K,2010DDA....41.0501N,2011MNRAS.417..198V,2013MNRAS.433.1813S,2015ApJ...803...80K,2019MNRAS.486.2995M,2020arXiv200700356R}. The presence of gaps and density perturbations within streams can inform the population of dark matter substructure, and subsequently the properties of dark matter \citep{2012ApJ...760...75C,2016MNRAS.457.3817S,2017MNRAS.470...60E,2019ApJ...880...38B,2019MNRAS.484.2009B,2020ApJ...892L..37B}. They can also be used to empirically track the underlying distribution of dark matter \citep{2012JCAP...08..027P,2019ApJ...883...27N}. 

Starting with the Sloan Digital Sky Survey (SDSS) \citep{2000AJ....120.1579Y}, numerous surveys  
have increased the number of cataloged stellar streams \citep{2001ApJ...548L.165O,2002ApJ...569..245N,2006ApJ...645L..37G,2006ApJ...642L.137B,2018ApJ...862..114S}. 
Most recently, the {\it Gaia} Space Telescope \citep{2018A&A...616A...1G,2018A&A...616A...2L} has opened a new frontier of Galactic kinematics and thus new opportunities for the discovery and study of stellar streams. 

Numerous successful stream-finding techniques have been applied to the {\it Gaia} data \citep{2018MNRAS.477.4063M,2018MNRAS.478.3862M,2018ApJ...863...26Y,2019A&A...621L...3M,2020MNRAS.492.1370B,2019A&A...622L..13M,2020arXiv201205245I}. In some cases cross-referencing \Gaia with other spectroscopic catalogs can provide additional kinematic or spectroscopic information, although statistically limiting the sample size (see e.g. \textsc{StarGo} \citep{2018ApJ...863...26Y}, which identifies streams in the cross match of {\it Gaia} DR2 with LAMOST DR5 \citep{2015RAA....15.1095L}). Of the methods relying exclusively on {\it Gaia}, the \textsc{Streamfinder} algorithm \citep{2018MNRAS.477.4063M,2018MNRAS.478.3862M} leverages the fact that stars within a stellar stream would have similar orbits through the Galaxy. By searching for stars occupying the same ``hypertubes'' through six-dimensional position/velocity space, \textsc{Streamfinder} has discovered a number of new stellar streams 
\citep{2018MNRAS.481.3442M,2019ApJ...872..152I,2019ApJ...886L...7M,2020arXiv201205245I}. In order to construct these orbits, \textsc{Streamfinder} must assume a form for the Galactic potential, and search for stars on an isochrone as part of a kinematically cold stream.

In this paper, we present \textsc{Via Machinae}, a new  algorithm for automated stellar stream searches with {\it Gaia} data. Based on unsupervised machine learning techniques, we identify streams as local overdensities in the angular position, proper motion, and photometric space of stars in \Gaia DR2. Importantly, \textit{we do not assume the stars in question lie on a particular orbit or stellar isochrone.} In fact, the initial (and most computationally-intensive) machine learning training steps of \textsc{Via Machinae} are designed to find \textit{all anomalous} structures first, in an agnostic manner. Only then do we implement selections based on prior knowledge of the properties of known stream candidates (particularly that the stars are distributed in an approximately linear structure over small angles on the sky). Such choices can be modified to target structures with other distributions in stellar photometry and proper motion, for example globular clusters or debris flow \citep{Lisanti:2011as,Kuhlen:2012fz}.\footnote{Debris flow refers to structure localized in velocity space, but incoherent in physical space \citep{1999MNRAS.307..495H,2012PDU.....1..155L,Kuhlen:2012fz}. This is usually the case for older mergers, e.g. the {\it Gaia} Sausage/Enceladus \citep{2019ApJ...874....3N}.} This flexibility may allow our technique to be sensitive to a wider variety of stellar streams than previous methods, and can be generalized to other anomalous features within the {\it Gaia} dataset (or other astrophysical surveys).

\textsc{Via Machinae} has two main components: an anomaly finding algorithm, and a line finding algorithm. The first component is the ANODE (ANOmaly detection with Density Estimation) algorithm (\cite{Nachman:2020lpy} hereafter referred to as \citetalias{Nachman:2020lpy}). 
Originally developed to search for new physics at the Large Hadron Collider, ANODE is a general machine learning algorithm for finding localized overdensities in any dataset. To accomplish this, ANODE leverages recent advances in density estimation using neural networks, specifically the idea of {\it normalizing flows} (for a recent review and original references, see e.g.~\cite{papamakarios2019normalizing}). In this paper, following the original ANODE work \citepalias{Nachman:2020lpy}, we use Masked Autoregressive Flows (MAF) \citep{papamakarios2018masked} to estimate the probability densities of stars in the \Gaia dataset.

The ANODE algorithm begins by slicing up the dataset into search regions and their complements, the control regions. As kinematically cold streams are expected to be fully localized in both proper motions, we choose to split the dataset into search regions consisting of slices in one of the proper motion coordinates. Then we use the MAF to estimate the probability distribution in position/proper motion/color/magnitude space of the stars in each search region in two different ways: (1) directly with the stars in the search region; and (2) indirectly with the stars in the control region, followed by interpolation into the search region. The interpolation step is a ``free" byproduct of the density estimation, because we actually learn a {\it conditional} probability density conditioned on the proper motion used to define the search region. If the search region contains a stream while the control region does not, then (2) can be thought of as a data-driven estimate of the probability density of the ``background" (i.e.\ non-stream) halo stars in the search region. Taking the ratio of these two density estimates forms a discriminant $R$, which is sensitive to anomalous overdensities (or underdensities) in the search region. By selecting the stars with the largest likelihood ratios, we can preferentially enhance the presence of stream stars vs.\ background stars in any given search region.

After performing such a selection in each search region, we are left with a much reduced set of stars spread across the sky. Only some of these stars will correspond to stellar streams. The rest may be other interesting structures (e.g.~globular clusters or debris flow) or spurious false positive fluctuations of the ANODE algorithm. This leads to the second major component of the \textsc{Via Machinae} algorithm: an automated method to search for linear features in a collection of stars in an angular patch of the sky. Simply fitting the stars to a line using (for example) least squares regression yields extremely unsatisfactory results, owing the presence of noise and outliers (i.e.~in a collection of stars, only a small fraction might belong to the stream). Instead, we have developed a method based on the Hough transform. This is an age-old machine learning technique that was originally developed for finding lines and edges in photographs \citep{Hough:1959qva,10.1145/361237.361242}, but which we adapt here to accomplish the same purpose in scatter plots.\footnote{The Hough transform has also been proposed for stellar stream identification in \citep{2019ApJ...883...87P,pearson2021hough} in the context of M31.} 
The idea of the Hough transform is to convert the problem of line finding to
counting intersections of curves in an auxiliary parameter space (the Hough space). In this way, one can also give a (rough) figure-of-merit to the best-fit line detection, based on the contrast between regions of high and low curve density in Hough space.

\begin{figure*}
\begin{centering}
\includegraphics[width=1.9\columnwidth]{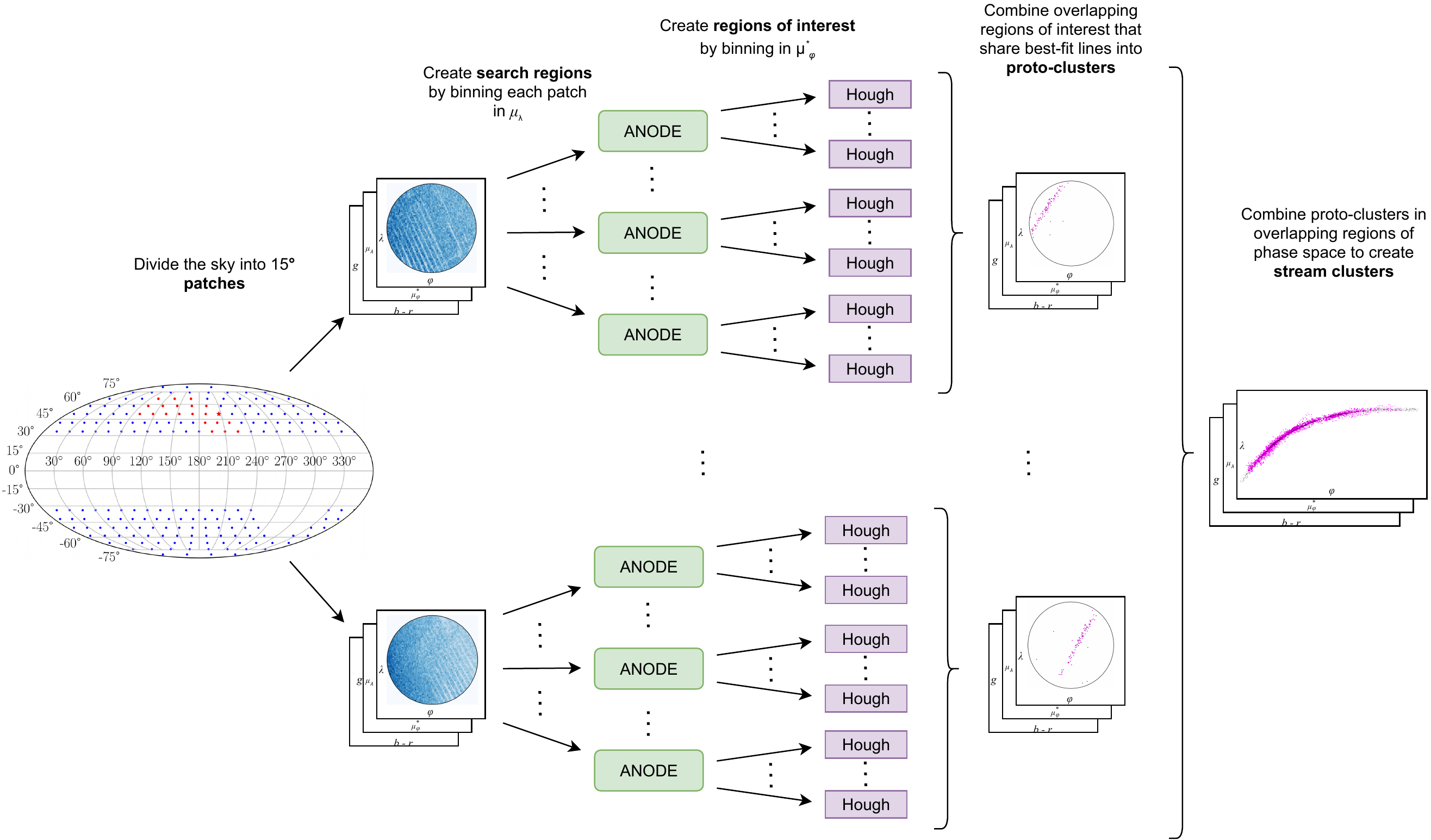}
\caption{A schematic showing an overview of the \textsc{Via Machinae} algorithm. Bolded and boxed terms are defined in \Sec{sec:algorithm} (with the exception of {\it patches}, which are described in \Sec{sec:inputs}). First we divide up the sky into evenly-tiled 15$^\circ$ patches. Within each patch, we further divide up the stars into search regions defined by a window in $\mu_\lambda$, one of the proper motion coordinates (the remaining data features for each star are denoted $\vec{x}$). Then we train the ANODE algorithm on the search regions and their complements, to learn a data-driven measure of local overdensities $R(\vec{x})$. To turn this measure into a stream finder, we further divide up the SRs into regions of interest based on the orthogonal proper motion coordinate $\mu_{\phi}^*$. We apply an automated line-finding algorithm based on the Hough transform to the 100 highest-$R$ stars in each ROI. Finally, we combine  ROIs adjacent in proper motion that have concordant best-fit line parameters into proto-clusters, and cluster these across adjacent patches of the sky into stream candidates.}
\label{fig:viamachinae_schematic}
\end{centering}
\end{figure*}

The major steps and key terms of \textsc{Via Machinae} are summarized in \Fig{fig:viamachinae_schematic}.
 Moving from left to right in this figure:
\begin{itemize}
    \item We divide the sky into overlapping {\it patches} of stars, each a circular region of radius $15^\circ$.
    
    \item These patches are then divided into overlapping {\it search regions} based on one proper motion coordinate. The estimated probability ratio $R$ for each star in each search region is obtained by ANODE training. We then limit ourselves to the inner $10^\circ$ of the patch to avoid edge effects (among other fiducial cuts).
    
    \item Each search region is then subdivided into {\it regions of interest} using the orthogonal proper motion coordinate which was not used to define the search region. In order to further purify signal to noise, a cut on color is imposed to focus on old, metal-poor stars that comprise the majority of known streams. 
    
    \item The 100 stars with the highest $R$ values in each region of interest are mapped to Hough space and the most line-like feature is assigned a significance $\sigma_L$.
    
    \item In overlapping regions of interest, we combine coincident lines and $\sigma_L$ values to obtain a proto-cluster for the patch, with an accompanying total significance $\sigma_L^{\rm tot}$.
    
    \item Proto-clusters in neighboring patches are combined into a stream candidate.
    
\end{itemize}

In this paper, we will use the GD-1 stream to illustrate the steps of the \textsc{Via Machinae} algorithm. GD-1 \citep{2006ApJ...643L..17G}, is an exceptionally long and dense stellar stream located at $\sim 10$ kpc, most likely originating as a globular cluster of mass $\sim 2 \times 10^4 M_{\odot}$ \citep{2010ApJ...712..260K}. When first detected using SDSS, GD-1 was thought to span $\sim 60^\circ$ in the sky. Using the second data release of \Gaia (\Gaia DR2), it has been extended by as much as 20$^\circ$ \citep{2018ApJ...863L..20P} (hereafter \citetalias{2018ApJ...863L..20P}), and was found to include gaps that could be evidence for dark matter substructure \citep{2018ApJ...863L..20P,2019ApJ...880...38B,2019arXiv191102663B,2021MNRAS.501..179M,2021MNRAS.tmp..242B}. Though most stellar streams are not nearly as long, dense, narrow, or well-defined as GD-1, it nevertheless provides an excellent testbed for \textsc{Via Machinae}, as stellar membership of the stream has been extensively studied (see e.g. \cite{2018ApJ...863L..20P,2019ApJ...880...38B,2020ApJ...892L..37B}), and its distinctiveness allows for clear demonstrations of the utility of the algorithm.

This paper is organized as follows: In \Sec{sec:inputs}, we introduce the {\it Gaia} data and its processing into inputs that will be used for anomaly detection. We then present the algorithm in \Sec{sec:algorithm}, with each step illustrated by its action on a segment of the GD-1 stream. In \Sec{sec:gd1}, we apply \textsc{Via Machinae} to the entire length of the GD-1 stream. Finally, in \Sec{sec:conclusions} we conclude with a summary and a list of interesting future directions motivated by this work. In a subsequent work \citep{full_sky}, hereafter \citetalias{full_sky}, we will apply our technique across the full \Gaia DR2 dataset, and demonstrate its ability to detect other known streams, and present new stream candidates.  

\section{Data and Input Variables}
\label{sec:inputs}

Before introducing the \textsc{Via Machinae} algorithm, we must first describe the data upon which it will be applied, and the pre-processing required. 

Starting with the \Gaia DR2 dataset,\footnote{As this work was being completed, \Gaia EDR3 \citep{2021A&A...649A...1G} was released. While our results  likely would have been improved by using this new dataset, re-running the ANODE method on \Gaia EDR3 proved to be too computationally expensive (the full-sky scan of \Gaia DR2 took ${\mathcal O}(10^5)$ NERSC-hours). We plan to apply our method to \Gaia EDR3 in a future publication.} we limit ourselves to distant stars with measured parallax less than 1~mas (corresponding to stars beyond 1~kpc). We do not correct for the {\it Gaia} DR2 zero-point parallax offset; varying the parallax cut by $\pm 0.05$ results in only a $\sim 3\%$ change in the number of stars and so is highly unlikely to affect our algorithm. We tile the sky with 15$^\circ$ patches using \textsc{Healpy} \citep{2005ApJ...622..759G,Zonca2019} (with $\rm{nside}=5$). This patch size was selected to have a tractable number of stars for the machine learning training step of the algorithm, as will be described in \Sec{sec:anode}. The patches are also large enough to capture significant portions of most known streams if they should pass through them. As stars in the Galactic disc would overwhelm the training, we limit the analysis to high Galactic latitudes $|b| >30^\circ$. We also exclude all patches that overlap with the LMC or SMC. The final result is 200 patches in total. 

For stars within a patch, our data consists of two position, two kinematic, and two photometric parameters: the angular position on the sky (e.g., right ascension [ra, $\alpha$] and declination [dec, $\delta$]), the corresponding angular proper motions ($\mu_\alpha \cos \delta$ and $\mu_\delta$), the magnitude of the star in the {\it Gaia} $G$-band ($g$), and the difference in the $G_{BP}$ and $G_{RP}$ {\it Gaia} bands  ($b-r$). Throughout this work, we will not correct for dust or extinction; especially since we confine ourselves to high Galactic latitudes, these corrections are generally small ($\lesssim 0.1$ for $b-r$) and do not vary much across a patch. Since we are only interested in local overdensities in each patch and will select a wide range of color for our final analysis, dust and extinction corrections should not significantly affect our results.

The $(\alpha,\delta)$ coordinates do not have a Euclidean distance metric across the sky, and the resulting distortions across the patch, especially at high latitudes, could negatively affect our neural density estimation.\footnote{Density estimation on spheres and other non-Euclidean manifolds is an active area of research, see e.g. \cite{rezende2020normalizing}. We do not use these techniques in this work.} Therefore, for each patch (defined by a circle centered on $(\alpha,\delta) = (\alpha_0,\delta_0)$ in angular position), we rotate the positions and proper motions using \textsc{Astropy} \citep{astropy:2013,astropy:2018} into a new set of centered longitude and latitude coordinates $(\phi,\lambda)$ so that $ (\alpha_0,\delta_0) \to (0^\circ,0^\circ)$. The unit vectors for the rotated coordinate system, $(\hat{\phi},\hat{\lambda})$, are aligned with those of the previous unit vectors $(\hat{\alpha},\hat{\delta})$. Within each patch, we will calculate angular distances using a simple Euclidean metric in $(\phi,\lambda)$.
For notational simplicity, we will define the new proper motion coordinate $\mu_\phi\cos\lambda$ as $\mu_\phi^*$ for the remainder of the work (similarly $\mu_\alpha^* \equiv \mu_\alpha \cos\delta$).

The patches defined above will be used as input for the ANODE method, as will be described in \Sec{sec:algorithm}, using the features $(\phi,\lambda,\mu_\phi^*,\mu_\lambda,b-r,g)$.
After training ANODE on each patch, we impose a set of additional fiducial cuts on the data. As we will describe in more detail in \Sec{sec:anode}, these cuts are driven by the limitations of the MAF density estimator. Specifically, to avoid edge effects in the neural network output, the post-ANODE fiducial region studied in this paper is the inner $10^\circ$ of each patch with a magnitude cut of $g< 20.2$. Above this magnitude cut, the completeness drops rapidly \citep{2020MNRAS.497.4246B}; this choice also helps reduce (but does not completely eliminate) streaking in the data and other artifacts due to incomplete coverage of the dimmest stars in the {\it Gaia} DR2 data \citep{2018A&A...616A...1G}.

\begin{figure}
\includegraphics[width=0.9\columnwidth]{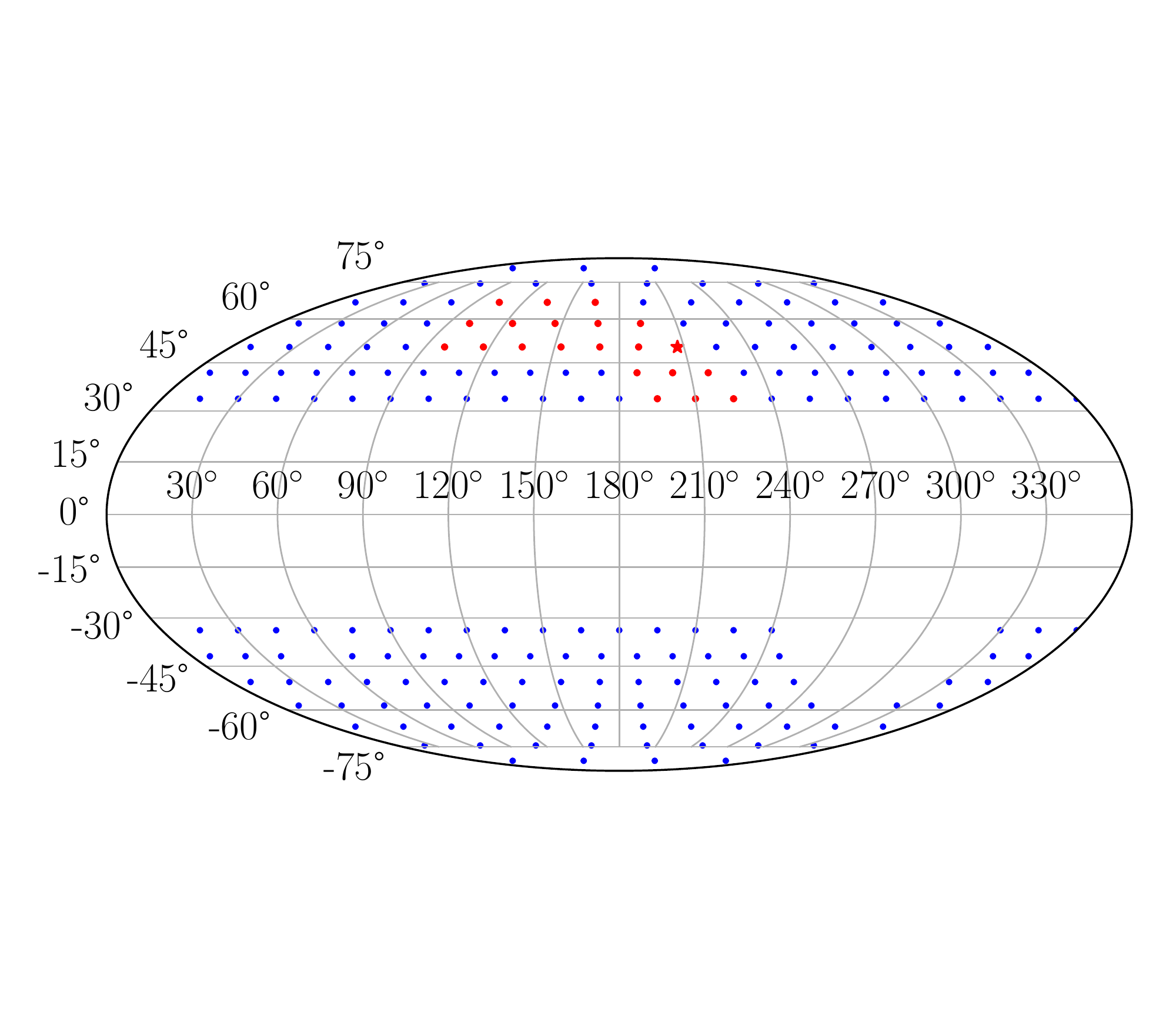}
\caption{The positions in Galactic $\ell$ and $b$ coordinates used for the centers for the datasets from the {\it Gaia} DR2 used in our full-sky analysis. The missing grid centers in the Galactic Southern hemisphere are the patches that overlapped with the Magellanic Clouds. The 21 centers which contain the GD-1 stream are shown in red, and the patch used as the worked example in \Sec{sec:algorithm} is denoted with a star.
  \label{fig:sky_locations}}
\end{figure}

As described in the Introduction, in this work we focus on demonstrating the \textsc{Via Machinae} algorithm using GD-1 as a worked example. Therefore, we limit ourselves here to patches of the sky that are known to contain portions of the GD-1 stream. We find that 21 patches in our all-sky sample include stars which have been identified by \citetalias{2018ApJ...863L..20P} as possible members of the GD-1 stream, for a total of 1,985 candidate GD-1 stars. Before (after) the ANODE fiducial cuts, the patches containing GD-1 have various numbers of stars, ranging from $8.0\times 10^5$ ($2.7\times 10^5$) in the patch with the least number of stars, to $2.1\times 10^6$ ($7.0\times 10^5$)  stars in the patch with the most number of stars.  \Fig{fig:sky_locations} shows the locations of all 200 patch centers we use to tile the sky as well as the 21 patches containing GD-1 stars. 

We will use the stream membership labels of \citetalias{2018ApJ...863L..20P} (which can be downloaded at \cite{price_whelan_adrian_m_2018_1295543}) as our point of comparison throughout this work. These were derived through relatively simple means: a visual inspection of the data, combined with polygonal cuts on proper motion, color and magnitude and a parallel strip cut (the ``stream track") in angular position. Thus we do not take them as ``absolute truth" labels -- indeed, some level of background contamination within this sample is certainly visible by eye.
Nevertheless, the GD-1 candidate labels of \citetalias{2018ApJ...863L..20P} still furnish a very useful and powerful point of comparison. 

\begin{figure*}
\begin{centering}
\includegraphics[width=1.9\columnwidth]{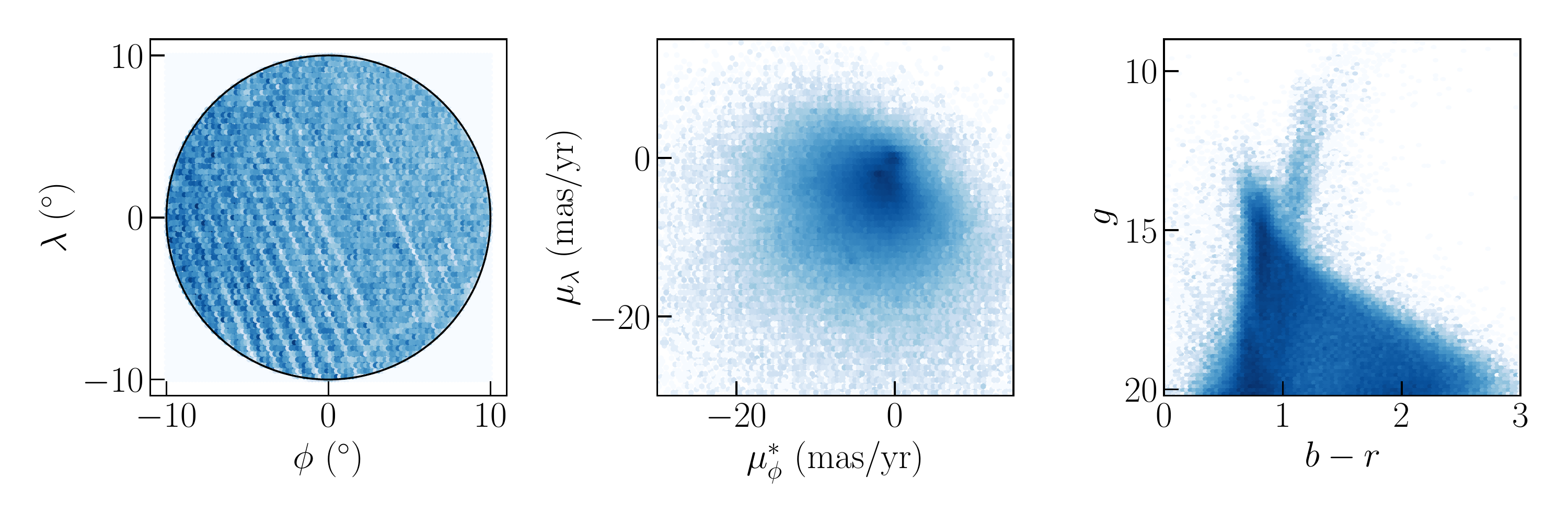}
\includegraphics[width=1.9\columnwidth]{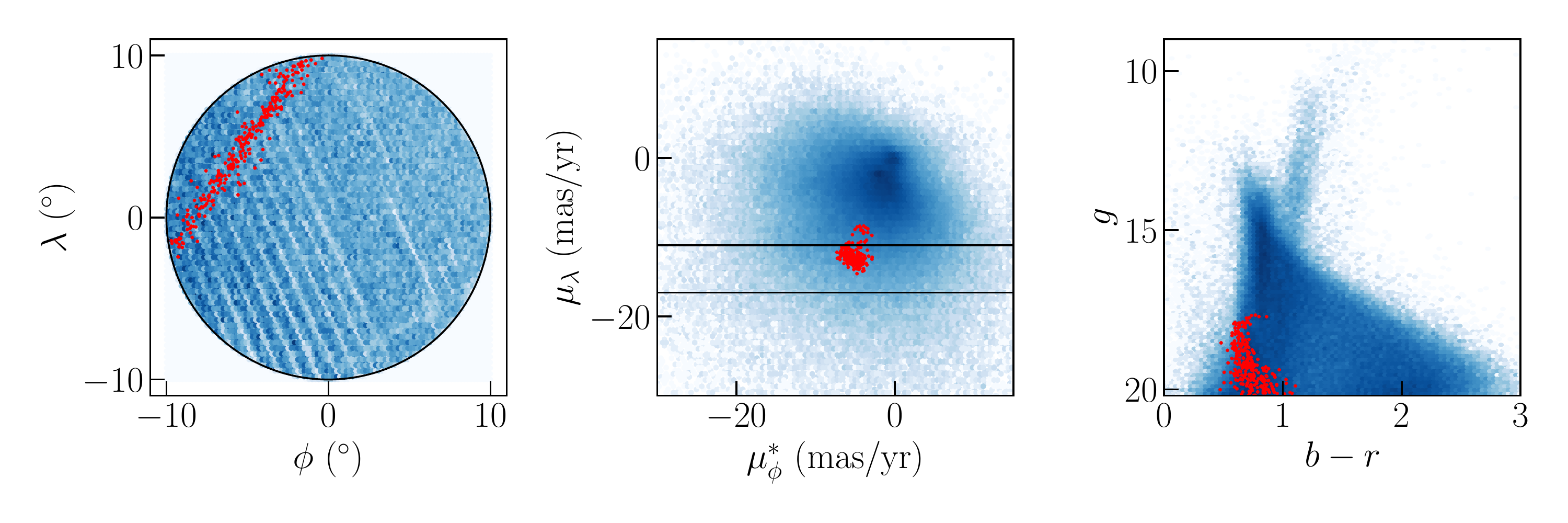}
\caption{Upper row: Angular position in $(\phi,\lambda)$ coordinates (left), proper motion in $(\mu_\phi^*,\mu_\lambda)$ coordinates (center), and photometry (right) of all stars in the patch centered on $(\alpha,\delta)=(148.6^\circ,\,24.2^\circ)$. (Note the streaking in angular position due to non-uniform coverage in \Gaia DR2.) Bottom row: As above, with stars identified by \citetalias{2018ApJ...863L..20P} as likely GD-1 stars shown in red, along with an example search region $\mu_\lambda \in [-17,-11]$~mas/yr in proper motion.  \label{fig:GD1example_allstars}}
\end{centering}
\end{figure*}

In \Sec{sec:algorithm}, we will use one of these 21 patches containing GD-1, centered on $(\alpha_0,\delta_0) =(148.6^\circ,24.2^\circ)$, to provide a worked example of each stage of \textsc{Via Machinae}. Within this patch's fiducial region, there are 334,376 stars, of which 276 have been identified as candidate members of GD-1 by \citetalias{2018ApJ...863L..20P}. The position, proper motion, and photometry of these stars is shown in \Fig{fig:GD1example_allstars}. In this patch, the candidate GD-1 stars lie in the range $\mu_\lambda \in [-14.6,-8.6]$~mas/yr.

\section{\textsc{Via Machinae}: The algorithm}
\label{sec:algorithm}

\subsection{ANODE: Defining the search regions}
\label{sec:SRs}

As described in the Introduction, the first part of \textsc{Via Machinae} is based on the ANODE method \citepalias{Nachman:2020lpy}. 
The starting point of ANODE is the subdivision of the stars within a single patch into {\it search regions} (SRs) which are windows in one feature of the dataset. The complement of the search region is called the {\it control region} (CR). The feature and the width of the window should be chosen so that, if a stream is present, there exists (at least) one SR which fully (or nearly fully) contains the entire stream. As we will explain in the next subsection, this is to enable accurate background estimation from the CR.
Defining the SRs by strips of angular position, for example, would not satisfy this requirement, unless the strips coincidentally aligned with the direction of the stream within the patch. However, stellar streams are kinematically cold and so are concentrated in both proper motion coordinates. Thus, we define our SRs using one of the proper motion coordinates. (Selecting SRs based on both proper motion coordinates is possible, but would greatly increase our total training time.) Since streams are localized in both proper motions, in principle it should not matter which one we choose; for this study, we choose $\mu_\lambda$ to be the proper motion coordinate defining the SRs.\footnote{The choice of proper motion coordinate can affect the performance of the algorithm through the number of background stars in the SR. For example, if the stream stars have small values of $\mu_\lambda$ but large values of $\mu_\phi$, then defining the SR in terms of $\mu_\lambda$ would lead to more background stars for the same number of stream stars, and hence a lower $S/B$, decreasing the stream detection probability. In \citetalias{full_sky} we will also incorporate the results of a scan over SRs defined using $\mu_\phi$ and show how this can achieve complementary results.} 

Based on the proper motion properties of known streams we find that a choice of a window in $\mu_\lambda$ of width 6~mas/yr is optimal. Streams like GD-1, located ${\cal O}(10~{\rm kpc})$ from the Earth, have proper motion dispersions of $\sim 2$~mas/yr. Such streams would be completely contained within our SRs at distances larger than $2-3$~kpc (which is more or less commensurate with the parallax cut we placed on our dataset). We also note that the stream does not have to be completely enclosed within a given SR for the algorithm to function. Proper functioning of ANODE requires only that the relative distribution of stars within the SR differ significantly from that in the CR; since the CR contains many more stars than the SR, a leakage of stream stars into the CR will not typically invalidate our approach. 

Since we do not know {\it a priori} which SR contains a stream, we must scan over all regions. In practice, we define a series of SRs by stepping in units of $\mu_\lambda=1$~mas/yr, with each SR then defined by the choice of $[\mu_\lambda^{\rm min},\mu_\lambda^{\rm max}]$:
\beq
[\mu_\lambda^{\rm min},\mu_\lambda^{\rm max}] = \ldots,[-10,-4],[-9,-3],\ldots,[3,9],[4,10],\ldots
\eeq
in units of mas/yr. The complement of the proper motion window (i.e.\ all the stars in the same patch that are not in the SR) defines the control region (CR) for each SR.

Each of these choices of $(\alpha_0 ,\delta_0, \mu_\lambda^{\rm min})$ furnishes a search region and control region pair for the ANODE training step. Overlapping the SRs in this way allows us to fully capture potential streams in at least one $\mu_\lambda$ window when performing a blind search -- if the SRs were not overlapping, then a stream could easily fall at the edge of two SRs, diluting the signal in each. By selecting SRs which are wide enough in proper motion to fully contain a kinematically cold stream and overlapping them by shifts which are smaller than the proper motion width of a typical stream, we minimize the possibility of this dilution. 

 SRs with fewer than 20k stars or more than 1M stars (before the fiducial cuts) are rejected for ANODE training. The former requirement is because too few stars in the SR results in poor density estimation performance, and the latter requirement is to avoid overly-long training times. In addition, SRs that contained a GC candidate (identified using a simple algorithm described in \App{app:GC}) were cut from the analysis, as the presence of the GC would completely overwhelm the training (i.e.\ in an SR containing a GC, the GC would correspond to such a large, delta-function-like overdensity, that ANODE would be unable to identify any other overdensity in the SR, such as one coming from a stream). In the end, we are left with a total of 545 SRs across the 21 patches of the sky containing GD-1.

To provide an example of an SR, we turn to our sample GD-1 patch defined in the previous section, centered on $(\alpha_0,\delta_0) =(148.6^\circ,24.2^\circ)$. We select the SR defined by $\mu_\lambda \in [-17,-11]$~mas/yr, which encompasses the majority of the GD-1 stars contained within this patch. This SR is shown in \Fig{fig:GD1example_allstars} and contains 34,823 stars in total, of which 252 are tagged by \citetalias{2018ApJ...863L..20P} as possible GD-1 members.

\subsection{ANODE: Density estimation}
\label{sec:anode}

Having defined the search regions, we turn to the probability density estimation step of the ANODE algorithm. As discussed in \Sec{sec:inputs}, the stars in our dataset are characterized by two position coordinates, two proper motion coordinates, color, and magnitude. Having set aside one of the proper motion coordinates $\mu_\lambda$ to define the search regions with, the remaining features $(\phi,\lambda,\mu_\phi^*,b-r,g)$ we will refer to collectively as $\vec x$. 

Suppose the stars in a patch consist of ``signal stars" coming from a cold stellar stream, and ``background stars" coming from the stellar halo. Let the conditional probability density of the background stars be $P_{\rm bg} (\vec{x}|\mu_\lambda)$, and the conditional density for the data (consisting of background stars plus signal stream stars) be $P_{\rm data}(\vec{x}|\mu_\lambda)=(1-\alpha) P_{\rm bg}(\vec{x}|\mu_\lambda)+\alpha P_{\rm sig}(\vec{x}|\mu_\lambda)$ where $\alpha$ is a measure of the signal strength. Then the optimal test statistic for distinguishing data from background is \citep{1933RSPTA.231..289N}:\footnote{Note that this will in general not be the optimal statistic for distinguishing any particular signal hypothesis from the background, rather it is the optimal test for distinguishing the background-only hypothesis from the data-driven probability distribution. For more discussion of the meaning of optimality in the context of anomaly detection, see the Appendix to \citetalias{Nachman:2020lpy}.}
\begin{equation}
R(\vec{x}|\mu_\lambda) = \frac{P_{\rm data}(\vec{x}|\mu_\lambda)}{P_{\rm bg}(\vec{x}|\mu_\lambda)}. \label{eq:Rdef}
\end{equation}
If the signal is small ($\alpha\ll1$) but sufficiently localized in feature space (i.e.\ a local overdensity), then we expect $R\gg 1$ where the signal is localized and $R\approx 1$ everywhere else. Since $R$ can be computed without knowing $\alpha$ or $P_{\rm sig}$,  selecting data points with high $R$ can purify signal to background in a model-agnostic way.

 Probability density estimation of arbitrary distributions is a difficult problem, and so  ANODE is only made feasible through recent advances in machine learning. In this paper, as in \citetalias{Nachman:2020lpy}, we employ the MAF architecture \citep{papamakarios2018masked} for the density estimation task. The MAF uses a specially-structured neural network to learn a bijective mapping from the original feature space into a latent space where the data is described by a unit multivariate normal distribution.\footnote{Our selection of hyperparameters is described in \App{app:hyperparam}.} 
 
Although it is relatively straightforward to train the MAF directly on the stars in the SR to learn $P_{\rm data}(\vec x|\mu_\lambda,\mu_\lambda\in {\rm SR})$ (the numerator of the likelihood ratio Eq.~\eqref{eq:Rdef}),  estimating the background density $P_{\rm bg}(\vec x|\mu_\lambda)$ takes more consideration. Calculating the denominator $P_{\rm bg}$ from first principles often proves impossible. Instead, one of the key ideas of the ANODE method is to use sideband interpolation from the CR (the complement of the SR) to estimate the background density in the SR. More precisely, we train a second MAF on the CR to learn $P_{\rm data}(\vec x|\mu_\lambda,\mu_\lambda\in {\rm CR})$. If there is no stream in the CR, then
\begin{equation}\label{eq:PdataCR}
P_{\rm data}(\vec x|\mu_\lambda,\mu_\lambda\in {\rm  CR})=P_{\rm bg}(\vec x|\mu_\lambda,\mu_\lambda\in {\rm CR}).
\end{equation}
If the background distribution in the CR is a smooth and slowly varying function of $\mu_\lambda$, then the MAF provides an automatic interpolation into the SR and yields an estimate for $P_{\rm bg}(\vec x|\mu_\lambda,\mu_\lambda\in {\rm SR})$, the denominator of Eq.~\eqref{eq:Rdef}.\footnote{If there is signal in the CR, then by assumption it will be a very small perturbation to $P_{\rm data}(\vec{x}|\mu_\lambda,\mu_\lambda\in {\rm CR})$ (i.e.~we assume there are many more background stars than signal stars in the CR). Then Eq.~\eqref{eq:PdataCR} will still be approximately true, and the signal contamination in the CR should not greatly affect the $R$ statistic in the SR.}

\begin{figure*}
\begin{centering}
\includegraphics[width=1.4\columnwidth]{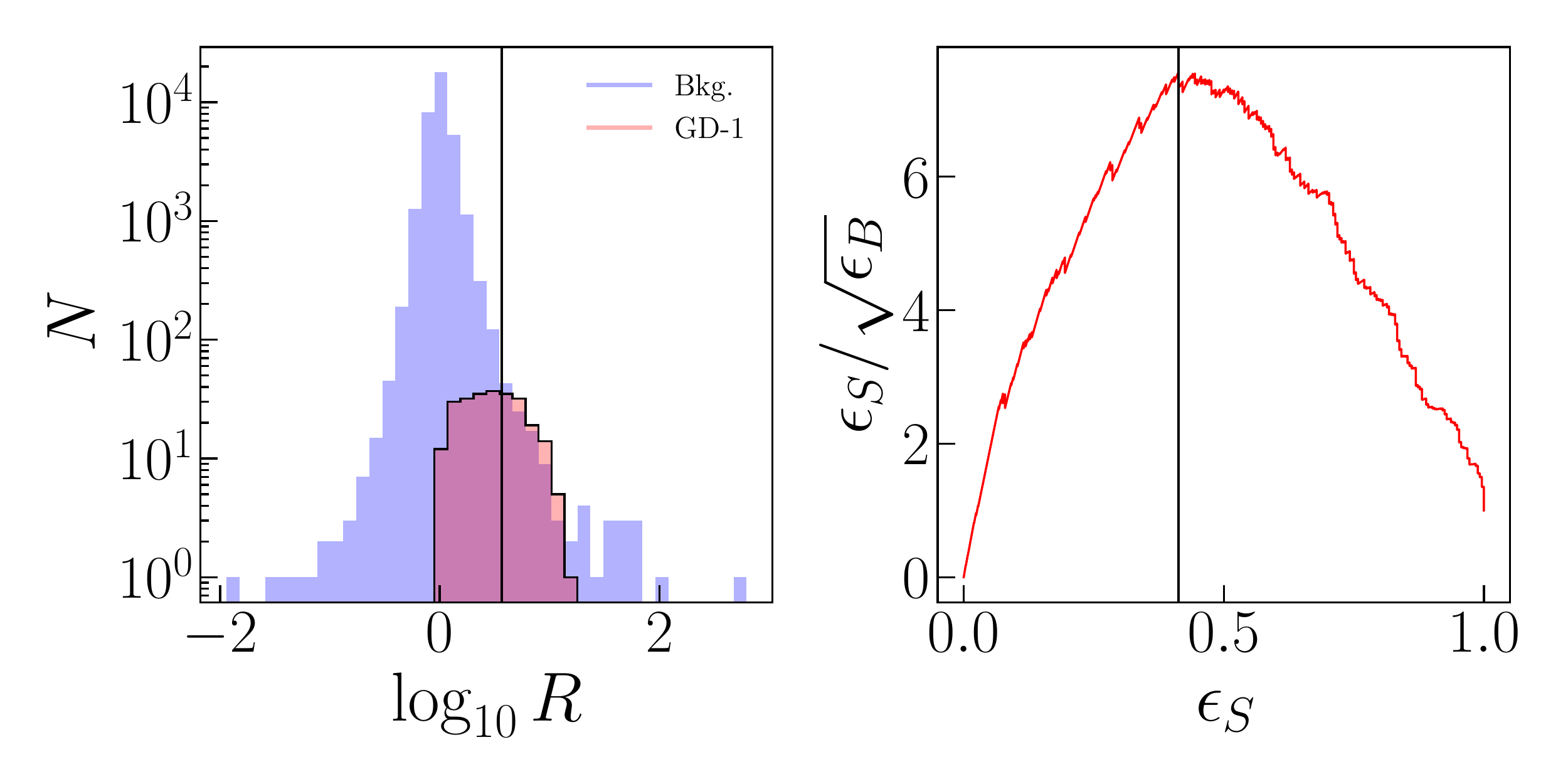}
\caption{Left: $R$ distribution for the SR $\mu_\lambda = [-17,-11]$~mas/yr in the patch centered at $(\alpha,\delta)=(148.6^\circ,\,24.2^\circ)$. Stars identified as likely members of GD-1 by \citetalias{2018ApJ...863L..20P} are shown in red, while the ``background" stars (those not tagged as likely GD-1 members by \citetalias{2018ApJ...863L..20P}) are in blue. Right: Significance Improvement Characteristic (SIC) curve for the same SR, showing the signal efficiency $\epsilon_S$ and the significance improvement (signal efficiency over square root of background efficiency, $\epsilon_S/\sqrt{\epsilon_B}$) as the cut on $R$ is varied. The vertical lines in both plots designate the $R$ value that maximizes the SIC curve.}
\label{fig:gd1_Randsic}
\end{centering}
\end{figure*}

\begin{figure*}
\begin{centering}
\includegraphics[width=1.9\columnwidth]{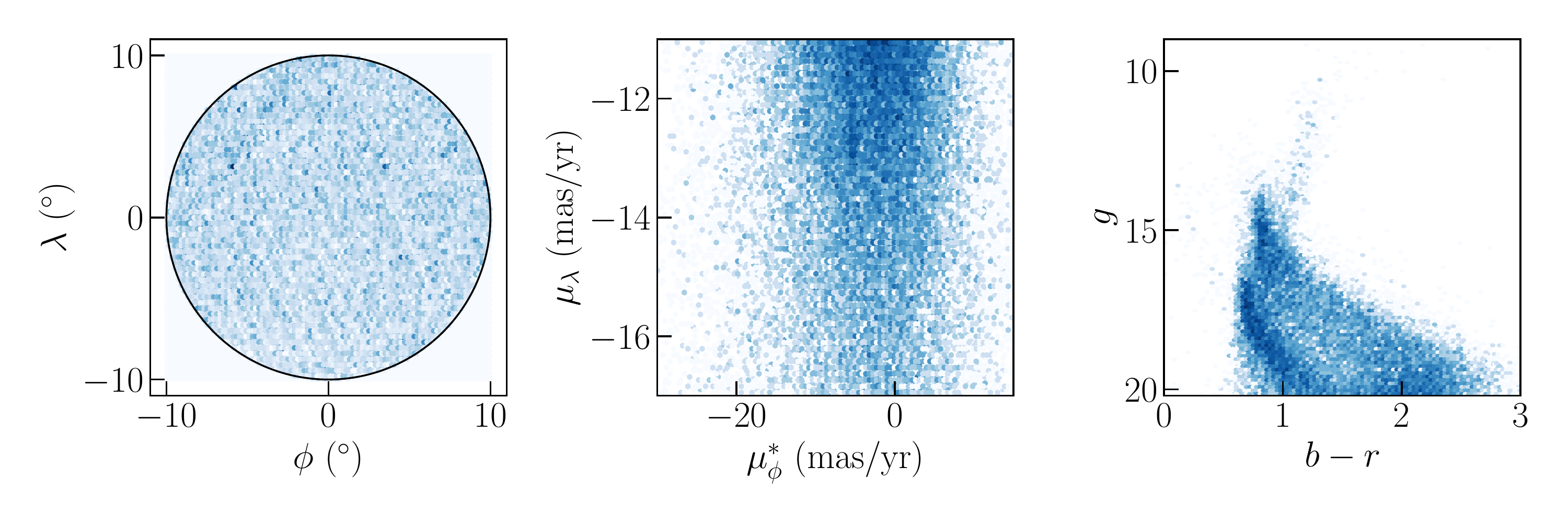}
\includegraphics[width=1.9\columnwidth]{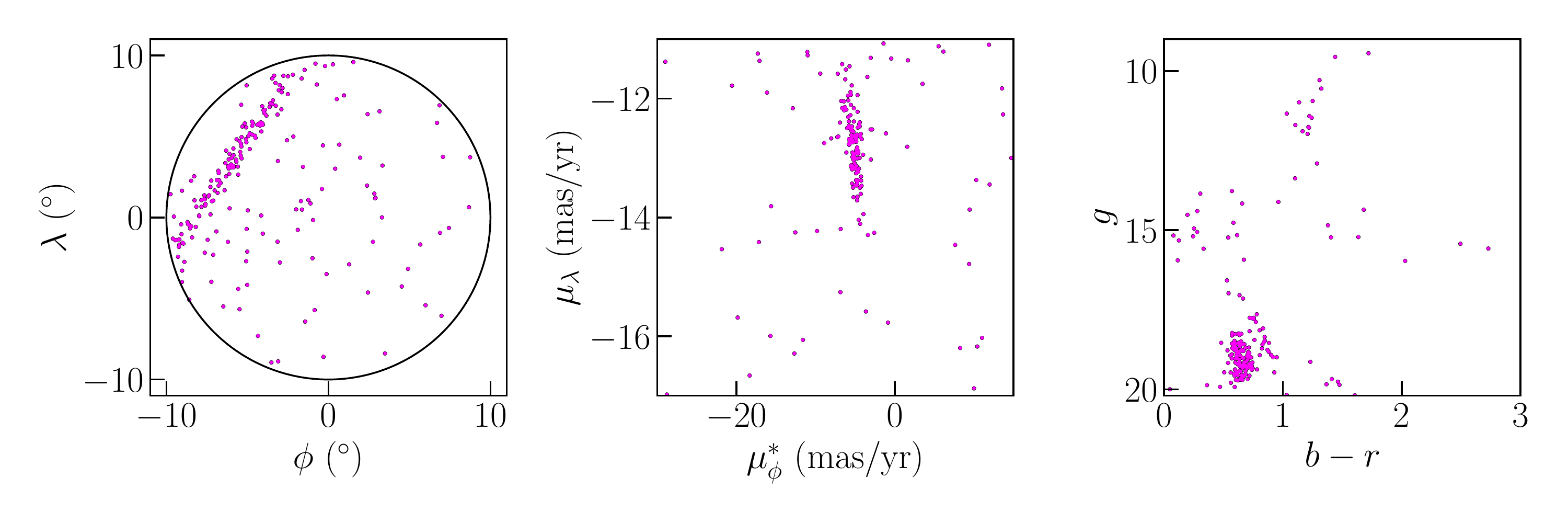}
\caption{Upper row: Angular position in $(\phi,\lambda)$ coordinates (left), proper motion in $(\mu_\phi^*,\mu_\lambda)$ coordinates (center), and photometry (right) of all stars (blue) in the $\mu_\lambda \in [-17,-11]$~mas/yr SR of our example patch centered on $(\alpha,\delta)=(148.6^\circ,\,24.2^\circ)$. 
Bottom row: As the upper row, applying the $R>R_{\rm cut}$ cut on the stars in the SR (purple). The GD-1 stream becomes immediately apparent. See text for details.  
}
\label{fig:GD1example_ANODE}
\end{centering}
\end{figure*}

An important point to note is that the MAF (along with most, if not all unsupervised density estimators) has difficulty matching rapid or discontinuous changes in the probability density as a function of the features $\vec{x}$. This is not a problem for the proper motion and $b-r$ features, which smoothly go to zero. However, in position-space, the selection of stars within a circular patch on the sky results in a sharp cutoff in density at the edge of the patch. Similarly, at high magnitude $g$, the sensitivity of the {\it Gaia} satellite drops rapidly. The result is spuriously large $R$ values near the edge of the patch in position space and at large $g$. To avoid this, we train on a larger dataset than the fiducial region in which we perform the subsequent stream-finding steps. As previously discussed in \Sec{sec:inputs}, after running ANODE, we define a fiducial region of $10^\circ$ around the center of the patch in $(\phi,\lambda)$ position space and a magnitude cut of $g< 20.2$.

In \Fig{fig:gd1_Randsic} (left), we show a histogram of the ANODE probability ratio $R$ for the stars in the $\mu_\lambda \in [-17,-11]$~mas/yr SR within the GD-1 example patch. We see that the likely GD-1 stars identified by  \citetalias{2018ApJ...863L..20P} are disproportionately represented at the high-$R$ tail of the ANODE distribution. By cutting on $R$, the resulting sample of stars would be enriched with stream stars compared to the full sample. For a given value of $R$, the signal efficiency $\epsilon_S$ is the fraction of candidate stream stars passing the cut on $R$, and the background efficiency $\epsilon_B$, is the fraction of non-stream-candidate stars passing the threshold. In \Fig{fig:gd1_Randsic} (right), we show the significance improvement characteristic (SIC) curve, comparing $\epsilon_S$ to $\epsilon_S/\sqrt{\epsilon_B}$ as $R$ is varied. We see that cutting on the ANODE output can greatly improve the purity of the sample and enhance the significance of the stream detection. For the sake of illustration, we have indicated in \Fig{fig:gd1_Randsic}  the optimal $R_{\rm cut}$ value, defined to be the cut on $R$ that maximizes the significance improvement in  \Fig{fig:gd1_Randsic} (right). (In more general settings, without stream-labeled stars, the optimal cut on $R$ would not be known, see the next subsection for further discussion of this.) Starting with 252 stars out of 34,823 identified as candidate GD-1 members by  \citetalias{2018ApJ...863L..20P},  the optimal $R_{\rm cut}$ value (corresponding to $\log_{10} R_{\rm cut} = 0.57$ for this SR) selects 206 stars, of which 103 are candidate GD-1 stars (corresponding to $\epsilon_S=0.41$). This nominally increases the statistical significance of the stream (i.e.\ $S/\sqrt{B}$) by more than a factor of 7. We emphasize that the $R$ ratio was learned in a completely data-driven, unsupervised manner, and at no point in the training were the stream candidate labels from \citetalias{2018ApJ...863L..20P} ever used. Here the labels are just used to illustrate the efficacy of the ANODE $R$-ratio in identifying stream stars.

In \Fig{fig:GD1example_ANODE} (top), we show all the stars in the $\mu_\lambda \in [-17,-11]$~mas/yr SR, and (bottom) those stars passing the optimal $R$ cut. GD-1 is an exceptionally dense and distinct stream: unlike other known streams it is visible, albeit barely, before the cut on $R$.  Performing the cut of $R>R_{\rm cut}$, as shown in the lower panel, drastically increases the significance of the stream, as expected.

Finally, we comment on the issue of streaking that can clearly be observed in the position space plots of the stars in many patches (\Fig{fig:GD1example_allstars} and \Fig{fig:GD1example_ANODE} are prime examples). These streaks are artifacts due to {\it Gaia}'s scan pattern and incomplete coverage of the sky in DR2. They might seem concerning for the ANODE method, as they appear as line-like overdensities in the angular coordinates, just like stellar streams would. However, we find no evidence that ANODE is incorrectly selecting for these spurious features. The reason is that ANODE looks for evidence of a {\it local} overdensity by comparing the stars in one proper motion slice with the stars outside of it. The streaking patterns are largely uncorrelated with proper motion; therefore, the overdensity they correspond to will actually {\it cancel} in the construction of the $R$ ratio, and these streaking stars will not be selected for by the ANODE algorithm.

\subsection{Regions of interest}
\label{sec:roi}

Up to this point, our method has been largely agnostic to the astrophysics of stellar streams (beyond the choice to use proper motion as our SR-defining feature). 
Stars tagged as anomalous by the ANODE training may be streams, globular clusters, debris flow, or some other structure localized in the Milky Way's velocity-space.
The steps in this and subsequent subsections are designed specifically to find cold stellar streams similar to the ones identified previously in data; different cuts and/or choices of parameters could be used to focus on other interesting astrophysical structures. The cuts we choose are:

\begin{itemize}

\item First, we remove all stars within a box around zero proper motion of width $2$~mas/yr. That is, we require 
\begin{equation}
    |\mu_{\lambda}|>2~\rm{mas/yr}~~ \rm{OR}~~ |\mu_\phi^*|>2~\rm{mas/yr}.
\end{equation}
Recall that the ANODE training identifies stars within the SR that are anomalous compared to the interpolation into the SR of the CR density estimate. Stars with proper motion near zero are predominantly distant stars; this population is not well-represented in a CR that does not contain $(\mu_\phi^*,\mu_\lambda) \sim (0,0)$~mas/yr. An example can be seen in \Fig{fig:pmcenter}.  If the SR contains this zero point, the distant stars are (correctly) identified as anomalous relative to the population in the control regions, but their sheer number completely overwhelms any other signal in the SR, requiring their removal after training is complete.

\begin{figure}
\includegraphics[width=0.9\columnwidth]{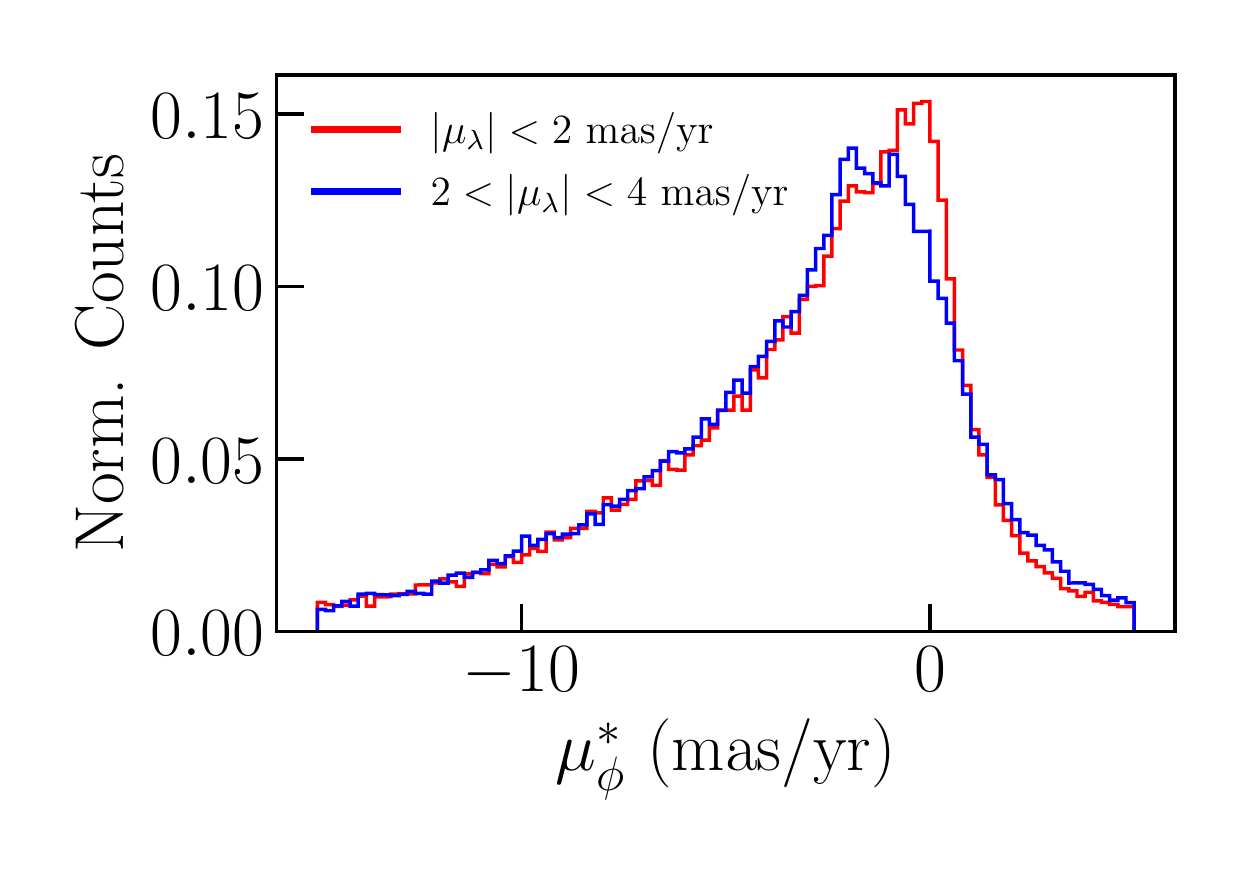}
\caption{Normalized histogram of $\mu_\phi^*$ values for stars in the $10^\circ$ patch centered on $(\alpha,\delta) = (148.6^\circ,24.2^\circ)$, requiring $2<|\mu_\lambda|<4$~mas/yr (blue) and $|\mu_\lambda|<2$~mas/yr (red). Note that the high density of stars near $\mu_\phi^* \sim 0$ with $|\mu_\lambda|<2$~mas/yr are not represented in the sample which does not overlap $\mu_\lambda \sim 0$. These very distant stars with near-zero total proper motion are absent as a population from search regions which do not include the zero point of proper motion.
 \label{fig:pmcenter} }
\end{figure}

\item Cold stellar streams, produced by tidally stripped globular clusters or dwarf galaxies, are predominantly composed of old, low metallicity stars. Many existing stream-finding algorithms leverage this by fitting stars in the stream candidate to isochrones appropriate to this assumption (see e.g. \cite{2018MNRAS.477.4063M}). Although the ANODE training is agnostic to such assumptions, in this work we are specifically interested in identifying cold streams, and not all anomalous overdensities. To that purpose, we now select stars in a specific color range in order to further purify signal to background. We require our stream candidates to lie in the broad range of colors $(b-r) \in [0.5,1]$. This range of colors was chosen so that it will contain (nearly) all of GD-1 and every stream found by  \textsc{Streamfinder} in \cite{2018MNRAS.481.3442M,2019ApJ...872..152I}. (\textsc{Streamfinder}  targeted globular cluster streams composed of stars with ages $\sim 10$~Gyr and metallicities -2~dex$\lesssim$ [Fe/H]$\lesssim$-1~dex). But being broader and more general than fitting to specific isochrones, we hope it will also enable the discovery of new streams. There may be interesting anomalous structures outside of this color range, which will be investigated in a future work.

\item To further isolate any potential streams, we subdivide the SRs defined by windows of  $\mu_\lambda$ into overlapping windows of $\mu_\phi^*$, with width 6~mas/yr and a stride of 1~mas/yr.
We call these windows {\it regions of interest} (ROIs) and they are labeled by $(\alpha_0,\delta_0,\mu_\lambda^{\rm min},\mu_\phi^{*\rm min})$. We exclude any ROI that has fewer than 200 stars as we need larger statistics to determine the presence of a stream.

\end{itemize}

Applying these cuts and further subdivision of the data to the 21 patches of the sky containing GD-1, we obtain 17,563 ROIs in total.

\begin{figure*}
\begin{centering}
\includegraphics[width=1.65\columnwidth]{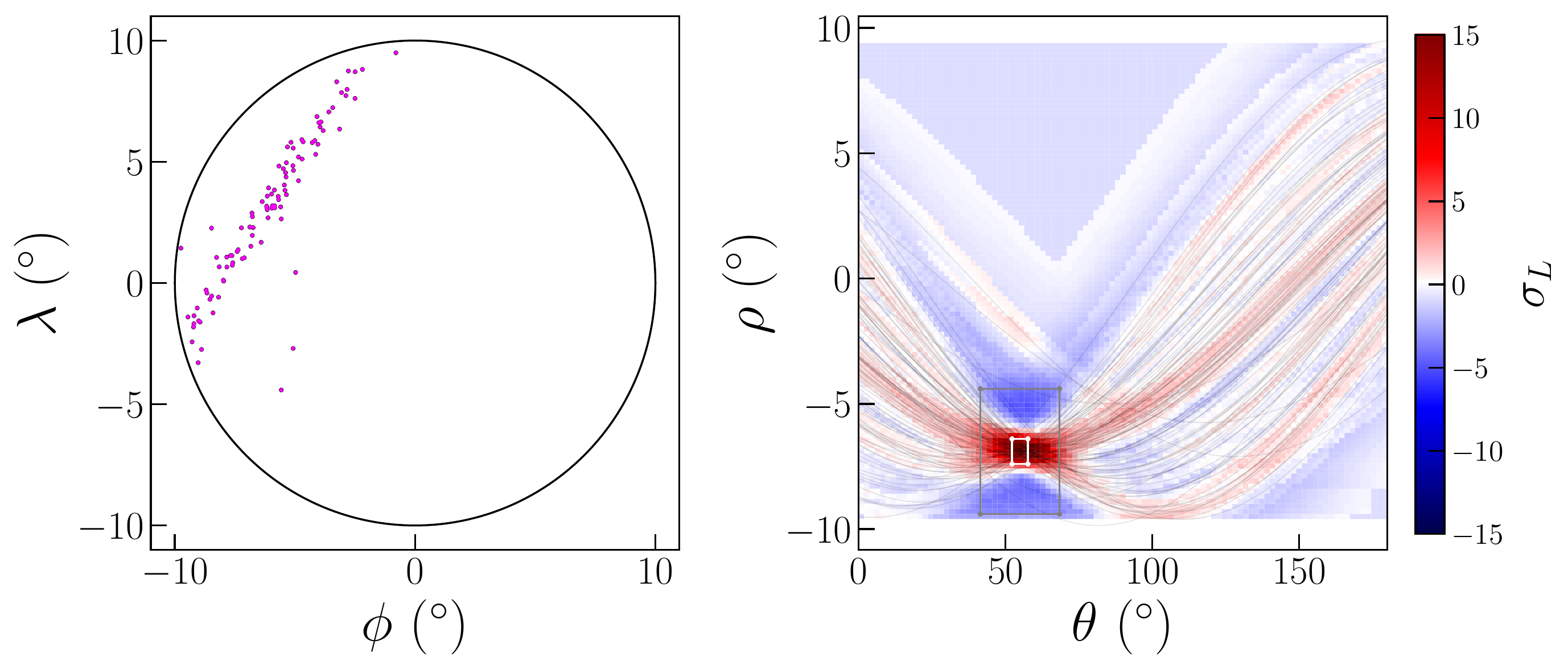}
\caption{Left: Angular position in $(\phi,\lambda)$ coordinates for the 100 highest-$R$ stars (purple) in the $\mu_\phi^* \in [-8,-2]$~mas/year, $\mu_\lambda \in [-17,-11]$~mas/year ROI from our example patch. Right: Associated curves in Hough space for these stars (black lines). The significance $\sigma_L(\theta,\rho)$ of a line oriented at each $(\theta,\rho)$ value is shown in color. The region around the point of maximum contrast (as identified by the \textsc{Via Machinae} algorithm) is indicated by the inner white box, with the region defining the background shown as the outer box.}
\label{fig:GD1example_SRhough}
\end{centering}
\end{figure*}

Within each ROI, we must decide how to apply the cut on the ANODE overdensity function $R(\vec x|\mu_\lambda)$. Many different types of cuts are possible, for instance setting a threshold as a percentile cut in each ROI, or a fixed value of $R$ across all ROIs. We have empirically found that selecting the 100 highest $R$ stars in each ROI is effective at finding known streams (more on this in \citetalias{full_sky}). An example of this is shown in the left panel of Fig.~\ref{fig:GD1example_SRhough}. 
It is possible that another cut (e.g.\ the 1000 highest $R$ stars in an ROI) would also be effective or would find other, qualitatively different streams. This would be interesting to explore in future work.

\subsection{Line-finding and stream detection}
\label{sec:linefinding}

Over large angles on the sky, most streams form arcs in $(\alpha,\delta)$ rather than lines (and streams with large line-of-sight velocities may not appear to form lines at all). However, the deviation from a line for the stars in the stream is small across a $10^\circ$ radius circle on the sky. 

Given the large number of ROIs -- ${\mathcal O}(10^4)$ for the 21 patches of the sky containing GD-1 alone -- we need an automated procedure for line finding. To do so, we adapt a long-standing technique from the field of computer vision based the Hough transform \citep{Hough:1959qva,10.1145/361237.361242}. A line passing through a point on the plane $(\phi,\lambda)$ can be expressed in terms of the distance $\rho$ of closest approach to the origin, and the angle $\theta$ between the $\phi$ axis and the perpendicular from the line to the origin:\footnote{Note $\rho$ can take negative values -- there is a periodicity in Hough space of the form
$(\rho,\theta)\sim (-\rho,\theta\pm\pi)$.}
\begin{equation}\label{eq:rhothetaxy}
\rho = \phi\sin\theta-\lambda\cos\theta.
\end{equation}
Viewing this equation another way leads to the idea of the Hough transform for line finding: for a single point, the collection of lines that pass through it will form a sinusoidal curve in the $(\theta,\rho)$ Hough space described by Eq.~\eqref{eq:rhothetaxy}.
If we consider two points in the plane, then their curves in Hough space will intersect for the values of $\theta$ and $\rho$ that define a line passing through both points. For a set of points in the plane, a subset of points on a line will manifest itself as overdensity in the $(\theta,\rho)$ space as many such curves intersect.

In \Fig{fig:GD1example_SRhough}, we show an example of the Hough transform on position data (left panel) of the 100 highest-$R$ stars in the ROI with $\mu_\phi^* \in [-8,-2]$~mas/yr, $\mu_\lambda \in [-17,-11]$~mas/year from our example patch. As can be seen in the right panel, the Hough curves for the stars on the line all cross at the same point, corresponding to the $\theta$ and $\rho$ values of the line on which the stream falls. The Hough transform therefore converts the problem of finding a line among a set of 2-dimensional points to the problem of finding the point with the highest density of curves in a 2-dimensional plane. Although this overdensity is obvious by eye in the example shown in \Fig{fig:GD1example_SRhough}, this is an extreme case and most overdensities will not be as clear-cut. 

We automate the line-finding by identifying the region in Hough space with the highest contrast in density compared to the region surrounding it.
We define a filter function which is applied to a box centered on a location $(\theta,\rho)$ of width $w_\theta$ and height $w_\rho$. The filter counts the number of stars whose Hough curves pass through the box, allowing us to define a number of curves at each point $n(\theta,\rho)$.  We then redo the filtering with a larger box (subtracting the curves which also pass through the initial box) to estimate the ``background'' curve count, $\bar{n}(\theta,\rho)$ (being careful to renormalize the counts for the different areas of the patch covered by the two regions in Hough space). Examples of these two filtering regions are shown in \Fig{fig:GD1example_SRhough}. The large and small box dimensions are ``hyperparameters" of the Hough transform line detection method and must be tuned based on known stellar streams to maximize detection efficiency. In this work, we will specialize to $w_\theta=5.4^\circ$ and $w_\rho=1^\circ$ for the inner box, and an outer box five times larger. This was found to be optimal for detecting relatively narrow streams such as GD-1. In \citetalias{full_sky} we will also explore other hyperparameters for the line finder that are sensitive to wider streams.

From the filter function count of Hough curves and background estimate at each point $(\theta,\rho)$, we define the line detection significance to be
\beq
\sigma_L(\theta,\rho) = {n(\theta,\rho)-\bar n(\theta,\rho)\over\sqrt{\bar n(\theta,\rho)}}
\eeq
We search in Hough space for the parameters that maximize this significance. Concretely, we bin the ($\theta-\rho$) plane in two dimensions, using a grid of 100 bins for $0 \leq \theta \leq \pi$ and 100 bins for $-10^\circ \leq \rho \leq 10^\circ$. We then select the bin that maximizes $\sigma_L$ and return this as our line detection in each ROI.\footnote{We are implicitly assuming here that each ROI will contain at most one stream. We believe this is a safe assumption, since ROIs are fully localized in both proper motions and angular position.}

\begin{figure}
\begin{centering}
\includegraphics[width=0.9\columnwidth]{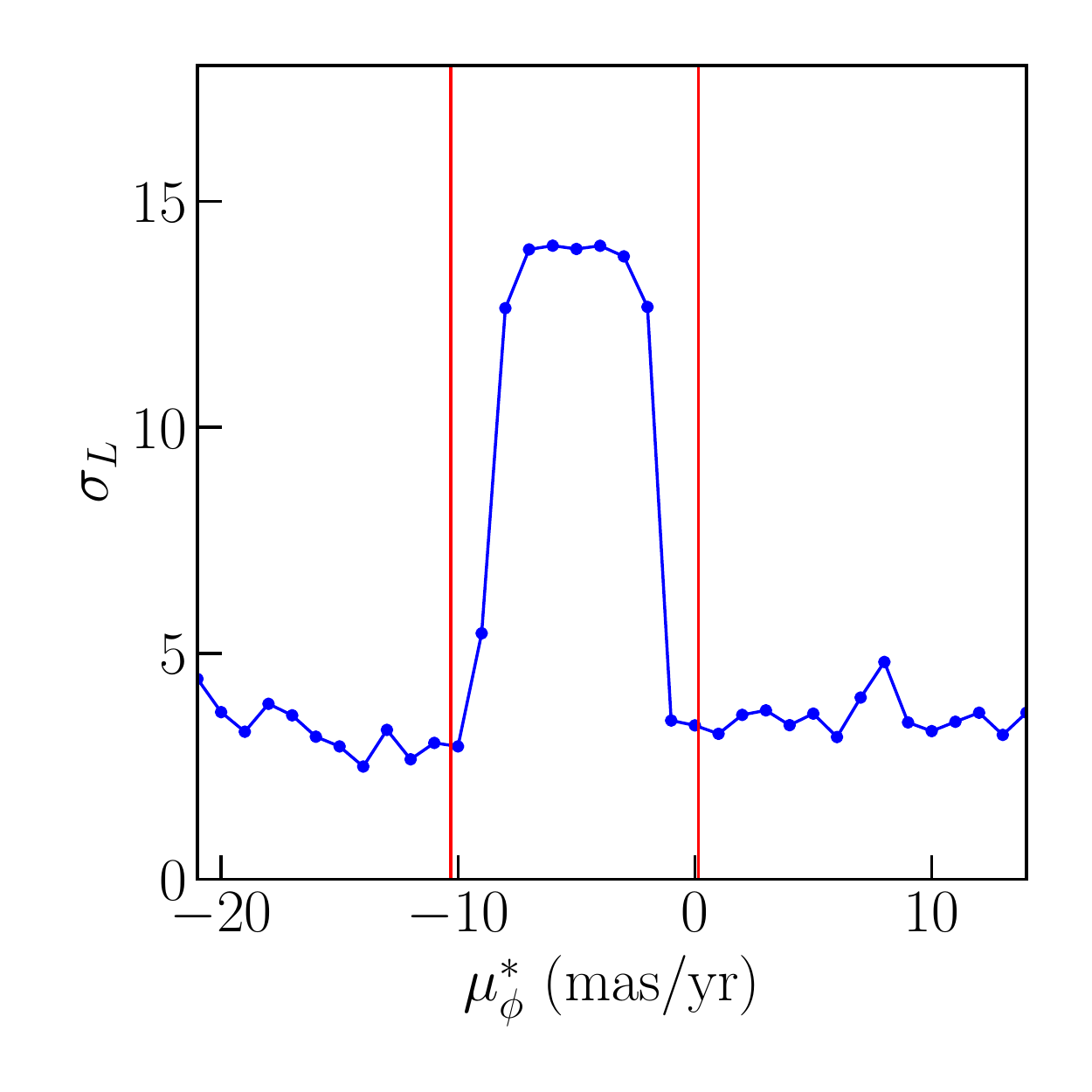}
\caption{Line significance $\sigma_L$ versus the central $\mu_\phi^{*}$ value for each ROI with $\mu_\lambda \in [-17,-11]$~mas/yr in our example patch. Vertical red lines indicate the minimum and maximum $\mu_\phi^*$ values for the candidate GD-1 stars of \citetalias{2018ApJ...863L..20P}.} 
\label{fig:gd1_ls}
\end{centering}
\end{figure}

When a stream is present in the SR and within the proper motion range of an ROI, we expect the resulting $\sigma_L$ value to be much larger than those of ROIs without linear structures. As an example of this, in \Fig{fig:gd1_ls} we show the $\sigma_L$ values for every ROI in the SR as a function of the central $\mu_\phi^*$ value defining each ROI, with vertical red lines indicating the maximum and minimum ROIs which contain any GD-1 stars. As can be seen, the high-significance lines fall only in the ROIs containing GD-1 stars. By cutting on $\sigma_L$, we are to be able to distinguish ROIs that contain an actual stream in the high-$R$ stars from those without.

\subsection{Final Merging and Clustering}
\label{sec:finalclustering}

After selecting the 100 highest $R$ stars in each ROI and applying the Hough transform line finder, we obtain the line parameters $(\theta,\rho)$ with the highest significance $\sigma_L$ in each ROI.  
We wish to use the significances of these lines to select only the most promising stream candidates. However, cutting on the raw $\sigma_L$ of an individual ROI is not effective in identifying a tractable number of likely stream candidates, because of the large trials factor (the so-called ``look elsewhere effect"). Across only the 21 patches containing GD-1 there are already ${\mathcal O}(10^4)$ ROIs, and random fluctuations could result in spurious line-like features in the background stars. This essentially dilutes the significance of a individual line detection by a correction factor, which may not be entirely trivial to estimate in the presence of correlations between ROIs. 

To obtain a meaningful line detection, we use the fact that a stream is likely to be found in multiple ROIs --
since the SRs are highly overlapping, each star generally has more than one $R$ value attached to it.
Therefore, we aim to cluster the ROIs that have concordant best-fit line parameters, across proper motions in a given patch, and across patches.

To perform this combination of overlapping ROIs, we have developed a three-step clustering algorithm (see Fig.~\ref{fig:viamachinae_schematic} for a graphical illustration of these steps):
\begin{enumerate}

\item In a given patch, we consider all ROIs with the same value of $\mu_\phi^*$. We group together ROIs adjacent in $\mu_\lambda$ which have concordant line parameters.\footnote{To be precise, we require the line parameters to be within $\Delta\theta=\pi/10$ and $\Delta \rho=2^\circ$ of each other.} In this way, all ROIs in a patch are clustered into {\it seeds} which have the same $\mu_\phi^*$ and consecutive values of $\mu_\lambda$. For each seed, we add the line significances of its ROIs in quadrature to form a combined line significance $\sigma_L^{\rm tot}$.\footnote{We are careful not to interpret $\sigma_L^{\rm tot}$ as a meaningful statistical significance in this work; rather we think of it more loosely as a figure of merit or an anomaly score for stream detection. At best, $\sigma_L^{\rm tot}$ would be a {\it local} significance (i.e.\ ignoring an enormous and difficult-to-quantify look-elsewhere-effect), and would be based on the assumption (probably not completely true) that separate ANODE runs in neighboring SRs return completely uncorrelated, random values of $R$ on background-only stars.}

\item Next, we group together seeds at adjacent $\mu_\phi^*$ based on the same criteria for concordance of line parameters. This forms {\it proto-clusters}, as shown in the second-to-last step of Fig.~\ref{fig:viamachinae_schematic}. 

\item Finally, we merge together proto-clusters across adjacent patches using the same criteria for concordance of line parameters. This produces our final {\it stream candidates}, as shown in the final step of Fig.~\ref{fig:viamachinae_schematic}.

\end{enumerate}

\begin{figure}
\begin{centering}
\includegraphics[width=0.9\columnwidth]{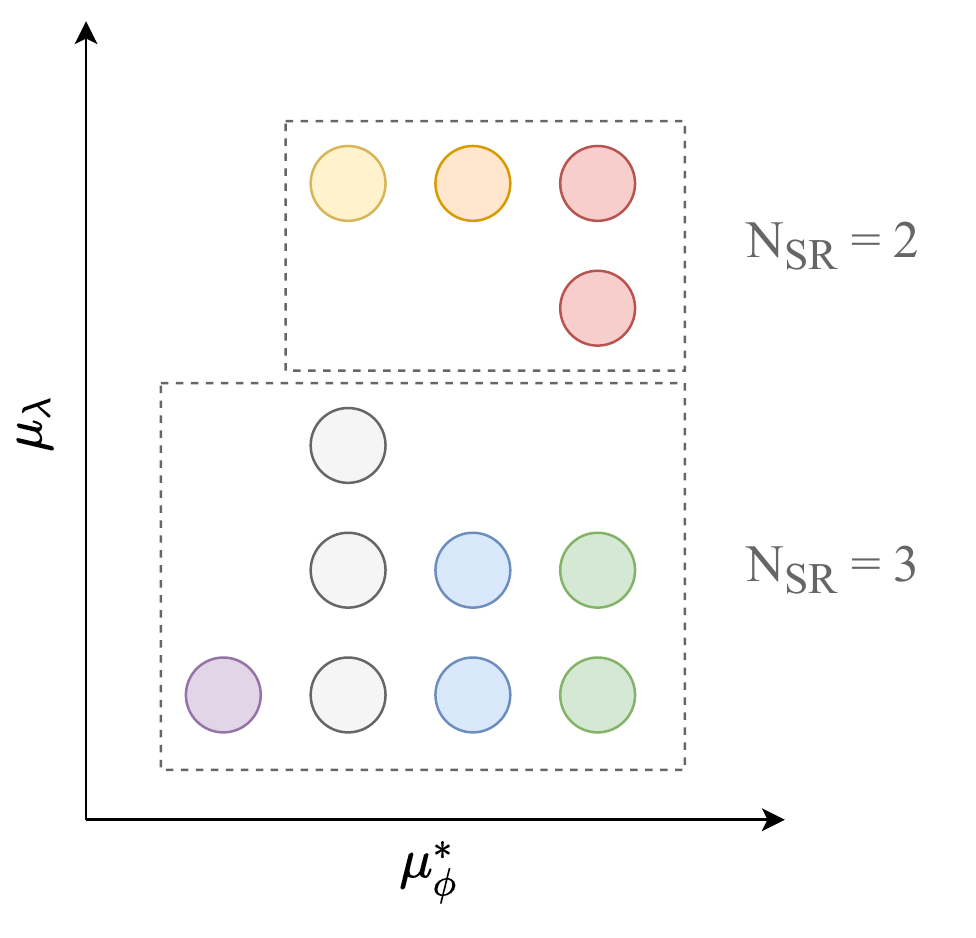}
\caption{A schematic showing how regions of interest (ROIs) are combined into different protoclusters. The different colors denote different seeds, i.e. clusters of ROIs with adjacent $\mu_\lambda$ and the same $\mu_\phi^*$ values. The boxes show how adjacent seeds are combined into protoclusters with different $N_{\rm SR}$.  } 
\label{fig:roi_schematic}
\end{centering}
\end{figure}

\begin{figure}
\begin{centering}
\includegraphics[width=0.9\columnwidth]{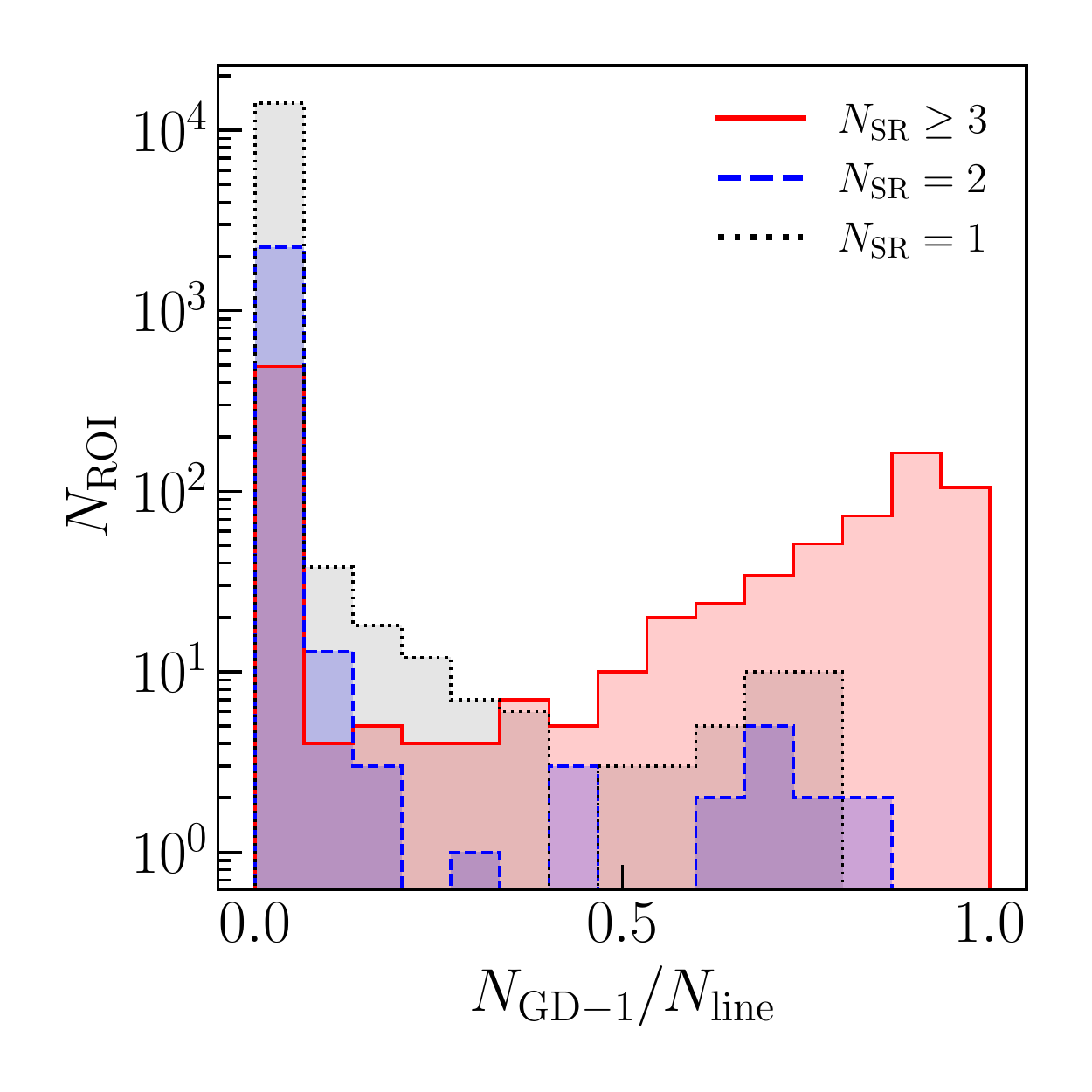}
\caption{Histograms of the fraction of
stars in the best-fit line of each ROI that were identified as likely GD-1 stars by \citetalias{2018ApJ...863L..20P}, for ROIs which are part of proto-clusters with $N_{\rm SR}=1$ (black, dashed), $N_{\rm SR}=2$ (blue, dotted) and $N_{\rm SR}\ge 3$ (red,solid). We see that requiring $N_{\rm SR}\ge 3$ greatly increases the fraction of candidate GD-1 stars in the best-fit line. } 
\label{fig:gd1_nroi_plot}
\end{centering}
\end{figure}

A schematic of steps (i - ii) is shown in \Fig{fig:roi_schematic}. There we see 12 hypothetical ROIs that are colored by seed. Neighboring seeds are combined into protoclusters which are denoted by dashed boxes. The number of signal regions, $N_{\rm SR}$, in each protocluster is the number of ROIs in its largest seed. 

In step (i), the rationale for grouping in $\mu_\lambda$ and not $\mu_\phi^*$ is that in a given patch, ROIs with the same $\mu_\lambda$ but different $\mu_\phi^*$ represent different, highly-overlapping slices of the same SR, with each star that appears in multiple ROIs having the same $R$ values from ANODE. On the other hand, ROIs with the same $\mu_\phi^*$ and different $\mu_\lambda$ represent different SRs, and each SR represents an independent ANODE training. Although the SRs are highly overlapping, the ANODE training is sufficiently stochastic that we take the outcome in different SRs to be quasi-independent. This motivates the adding in quadrature of the line significances of the ROIs in each seed. 

In step (ii), for each proto-cluster, we characterize its significance by the seed with the highest $\sigma_L^{\rm tot}$ that it contains. The size of this seed we will call $N_{\rm SR}$ and is another measure of the significance of the proto-cluster. Note that we do not add the $\sigma_L^{\rm tot}$ values of different seeds in a proto-cluster together in quadrature, since these are highly correlated.

Applying the final merging and clustering steps to the 21 patches containing GD-1, we find that the 17,563 ROIs are clustered into
10,955 proto-clusters. Of these, 10,267 have $N_{\rm SR}=1$; 
606 have $N_{\rm SR}=2$; and 82 have $N_{\rm SR}\ge 3$. 

All else being equal, we expect real streams to have higher values of $\sigma_L^{\rm tot}$ and $N_{\rm SR}$. 
We show in \Fig{fig:gd1_nroi_plot} histograms of the fraction of stars within the best-fit line of each ROI that have been identified as candidate GD-1 stars by \citetalias{2018ApJ...863L..20P}, for ROIs that belong with proto-clusters with different values of $N_{\rm SR}$. 
As can be seen, the fraction of candidate GD-1 stars (i.e.\ the ``purity" of the best-fit line) is significantly improved when we require $N_{\rm SR} \geq 3$. 

\begin{figure}
\begin{centering}
\includegraphics[width=0.9\columnwidth]{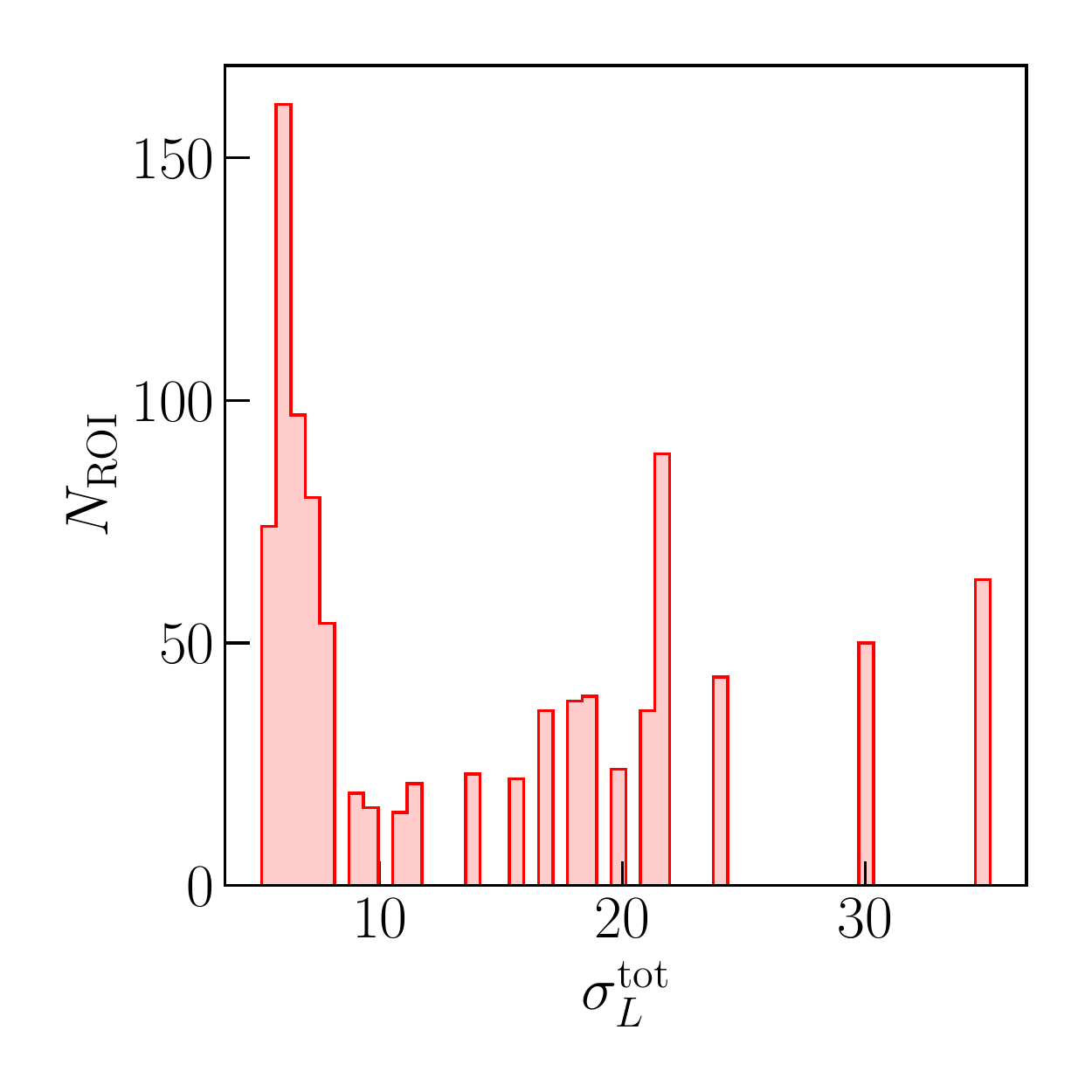}
\caption{Histogram of the $\sigma_L^{\rm tot}$ values of protoclusters with $N_{\rm SR}\ge 3$, with each protocluster weighted by the number of ROIs it contains.}
\label{fig:sigmahistNSR}
\end{centering}
\end{figure}

In \Fig{fig:sigmahistNSR}, we show the distributions of $\sigma_L^{\rm tot}$ values across the ROIs (the total line significance is for the proto-cluster that the ROI has been clustered into), for $N_{\rm SR}\geq 3$. We see that there is clearly a bulk distribution at low $\sigma_L^{\rm tot}$ and then a tail of outliers, with the separation occurring around $\sigma_L^{\rm tot} = 8$. It is reasonable to suppose that the majority of these low-$\sigma_L^{\rm tot}$ corresponds to false positives, while the tail could correspond to real stream detections that should be subjected to more in-depth investigation.

\section{Demonstrating the full Via Machinae Algorithm with GD-1}
\label{sec:gd1}

Having described all the steps of the Via Machinae algorithm, we now demonstrate the full algorithm on the 21 patches of the sky that contain GD-1. For the first step of the algorithm (ANODE), we used the ``Haswell" processors at NERSC, for a total of approximately 10,000 CPU-hours to analyze all 21 patches. For the subsequent steps of Via Machinae (line finding, forming protoclusters, and forming stream candidates), we used the local HEP cluster at Rutgers, for a total of approximately 50 CPU-hours.

\begin{figure}
\begin{centering}
\includegraphics[width=0.9\columnwidth]{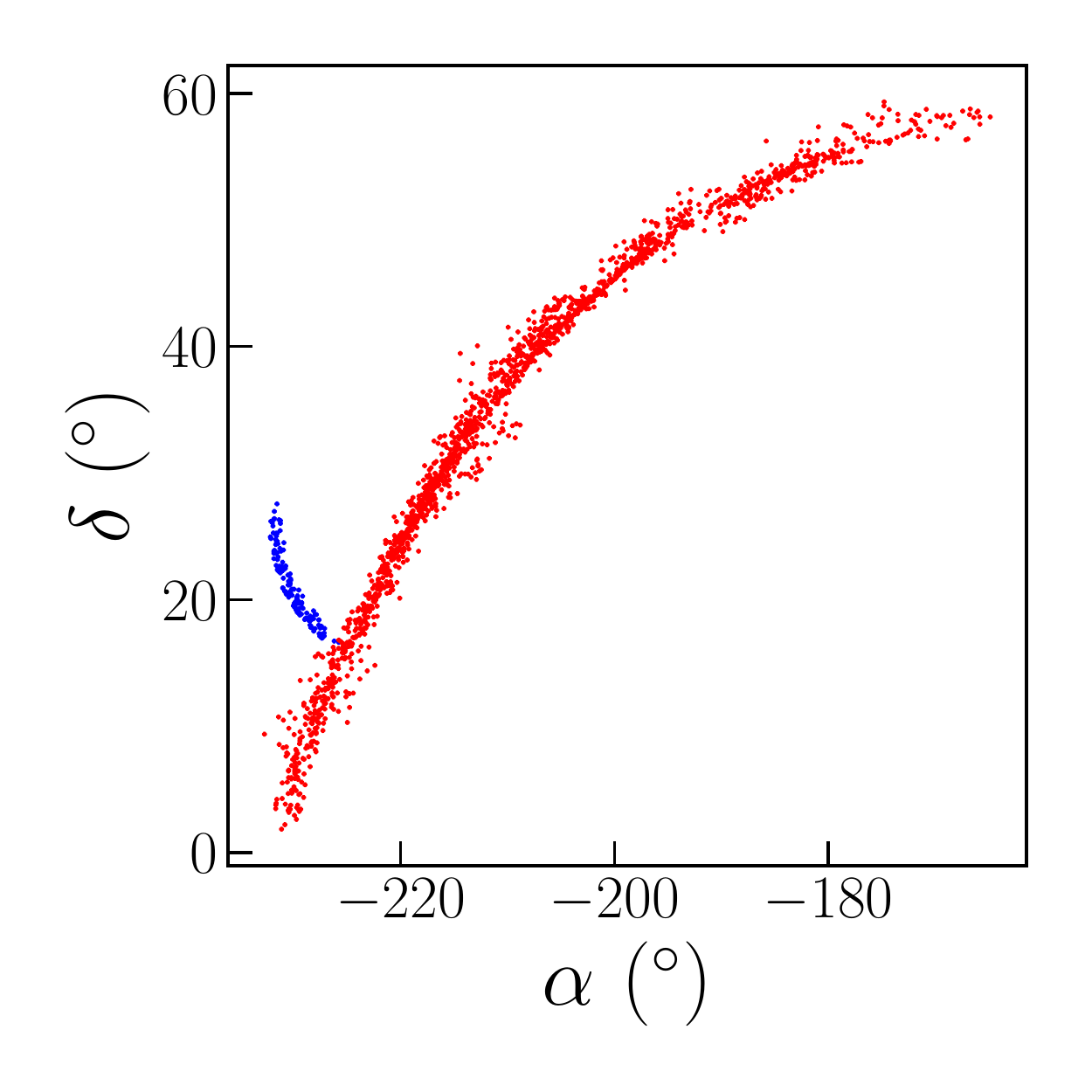}
\caption{The two stream candidates built out of proto-clusters with $N_{\rm SR}\ge 3$ and $\sigma^{\rm tot}_L \ge 7.5$. } 
\label{fig:highsigmastreams}
\end{centering}
\end{figure}

Motivated by the discussion in the previous subsection, we focus on only those proto-clusters with $N_{\rm SR}\ge 3$ and $\sigma_L^{\rm tot}\ge 8$. This leaves only 16 proto-clusters. Merging these results in  only two stream candidates, shown in \Fig{fig:highsigmastreams}. One might have expected far more stream candidates, given the enormous trials factors involved (e.g.  ${\mathcal O}(10^4)$ ROIs that we started with). This is a sign that the cuts on $N_{\rm SR}$ and $\sigma_L^{\rm tot}$ that we have chosen are indeed effective at reducing the false positive rate.

 The less prominent stream candidate, shown in blue, is built from a single proto-cluster representing 16 ROIs with $\sigma_L^{\rm tot} =9.5$. It comes from the patch centered at $(\alpha,\delta)=(138.8^\circ,\,25.1^\circ)$. The stream candidate does not correspond to any known stream, and {\it a priori} it may be a real stream or a spurious detection. Closer inspection reveals that all of the high-$R$ stars identified by ANODE are tightly clustered at the edge of the circular patch, almost perfectly aligned with the direction of the Galactic disk (and on the same side of the patch as the disk). Although this patch is $\gtrsim30^\circ$ off the Galactic plane, we still observe a strong density gradient towards and aligned with the disk. Therefore, we suspect that ANODE has identified disk stars in this case, and not a stellar stream. We discuss this further in \App{sec:stream2}.

\begin{figure*}
\begin{centering}
\includegraphics[width=1.9\columnwidth]{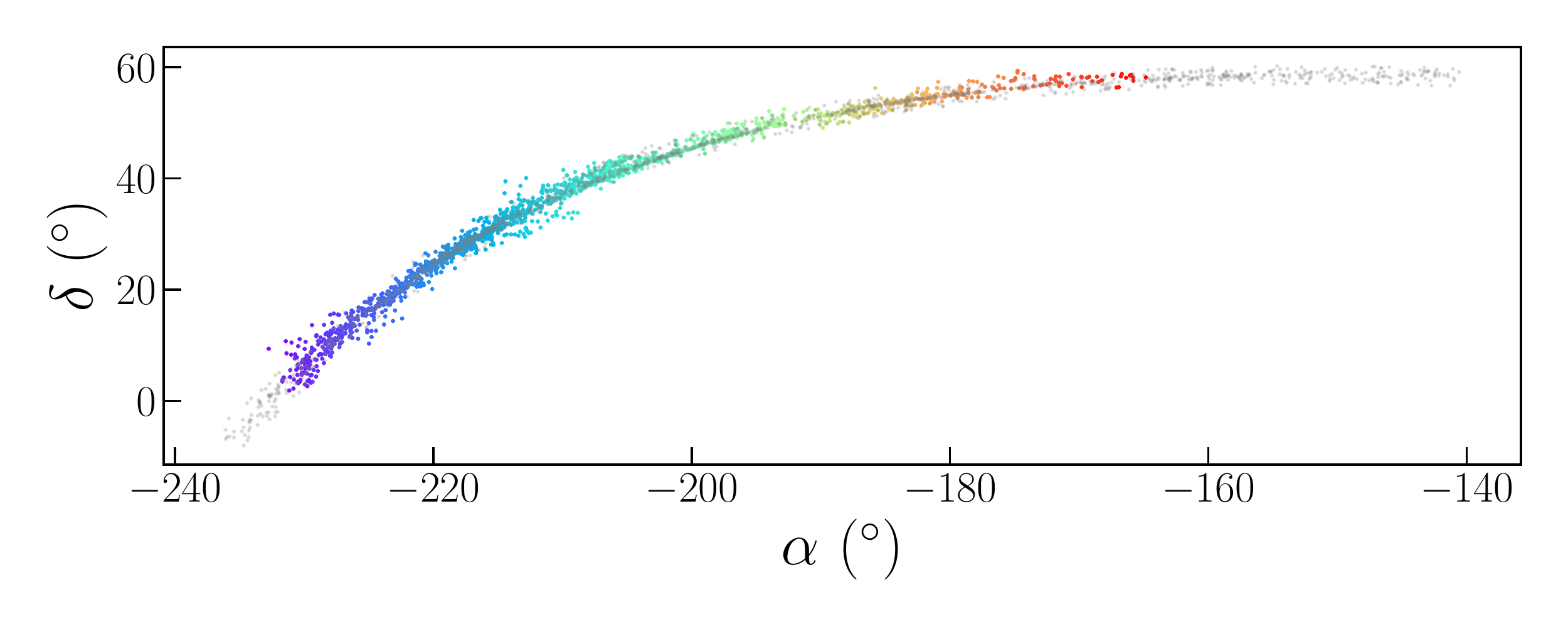}
\includegraphics[width=1.9\columnwidth]{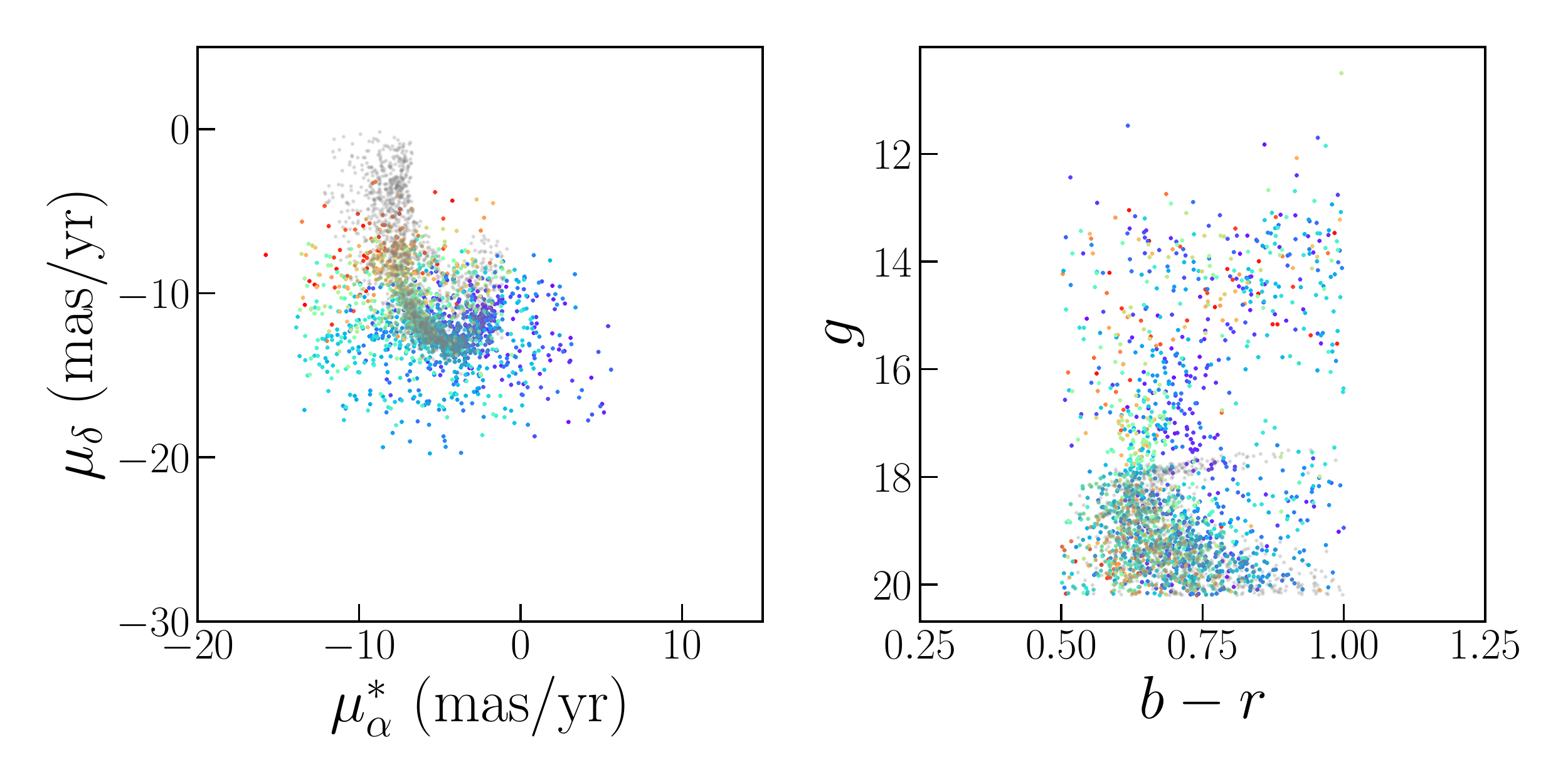}
\caption{Scatter plots of the angular positions, proper motions, and color/magnitudes of the 1,688 stars in the more prominent of the two stream candidates identified by \textsc{Via Machinae}, overlaid on the likely GD-1 stars tagged by \citetalias{2018ApJ...863L..20P} (gray) in the same region of $g$ and $b-r$ space. The \textsc{Via Machinae} stars are color-coded by position in $\alpha$, to facilitate cross referencing between the three individual scatter plots.}
\label{fig:gd1_finalcluster}
\end{centering}
\end{figure*}

\begin{figure*}
\begin{centering}
\includegraphics[width=1.9\columnwidth]{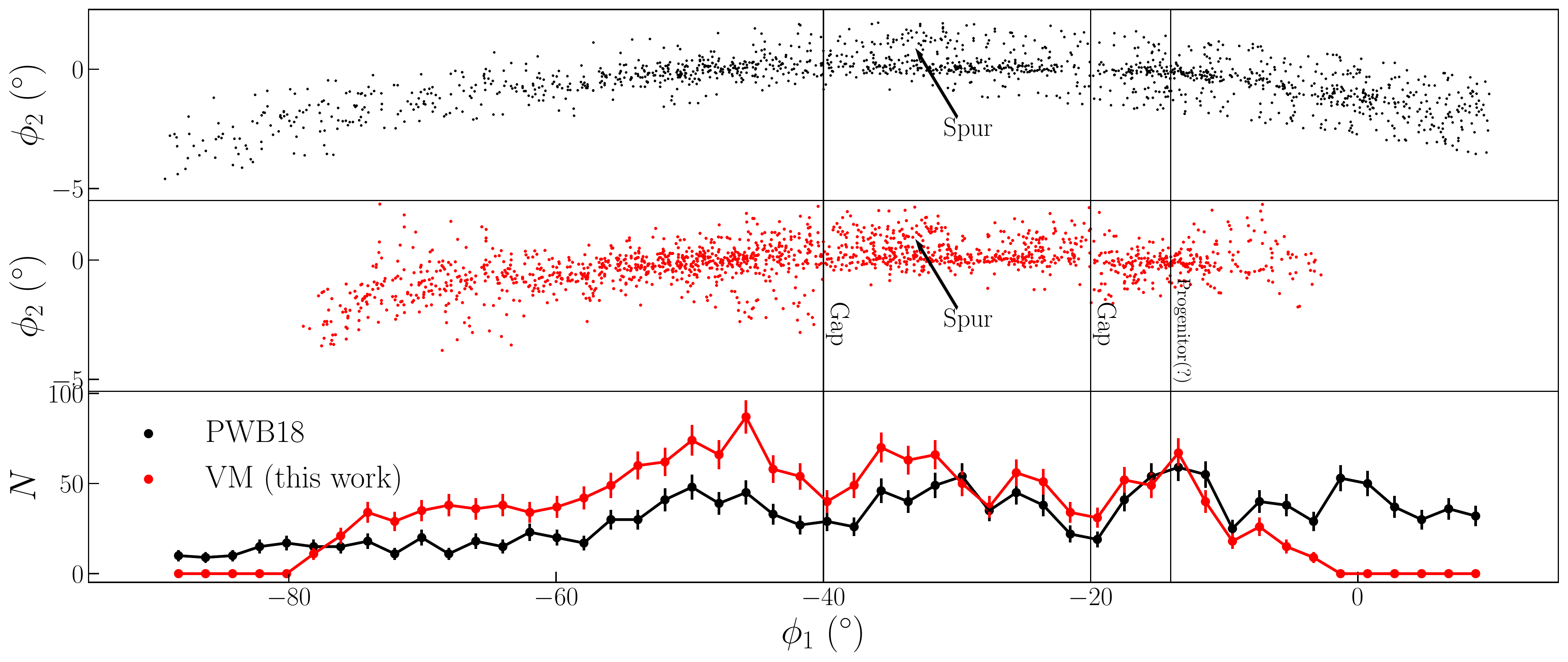}
\caption{Comparison of the likely GD-1 stars from \citetalias{2018ApJ...863L..20P} (top, black) and the stream candidate stars identified by \textsc{Via Machinae} (middle, red), in the GD-1 stream-aligned coordinate system $(\phi_1,\phi_2)$ \citep{2010ApJ...712..260K}. The location of previously identified features of GD-1 (two gaps, the possible progenitor location, and the spur) are indicated. Bottom row shows the number of candidate stream stars identified by \citetalias{2018ApJ...863L..20P} (black) and \textsc{Via Machinae} (red) in $\phi_1$ bins of width $2^\circ$; the error bars are purely statistical (Poissonian). In top and bottom panels, a cut on $g<20.2$ and $0.5<b-r<1$ has been applied to the stars from \citetalias{2018ApJ...863L..20P} so that a direct comparison can be made with the stars in this analysis.} 
\label{fig:gd1_streamcoords}
\end{centering}
\end{figure*}

The second, more prominent stream shown in red in \Fig{fig:highsigmastreams}, is composed of 15 proto-clusters representing 518 ROIs, and is clearly GD-1. 
 In \Fig{fig:gd1_finalcluster} we show the positions, proper motions, and photometry of the 1,688 stars in this stream candidate, overlaid on the locations of the stars tagged as likely GD-1 stream members by \citetalias{2018ApJ...863L..20P}. In \Fig{fig:gd1_streamcoords}, we present another look at the comparison between \textsc{Via Machinae} and \citetalias{2018ApJ...863L..20P}, this time using the coordinate system aligned with the GD-1 stream \citep{2010ApJ...712..260K}.\footnote{To allow for direct comparison of our results with  \citetalias{2018ApJ...863L..20P} in this section, we show the stars of the latter without correction for extinction, and apply the same cuts on their (uncorrected) magnitudes and colors of $g<20.2$ and $0.5<b-r<1$ as we do for our fiducial sample.}

\begin{figure}
\begin{centering}
\includegraphics[width=0.9\columnwidth]{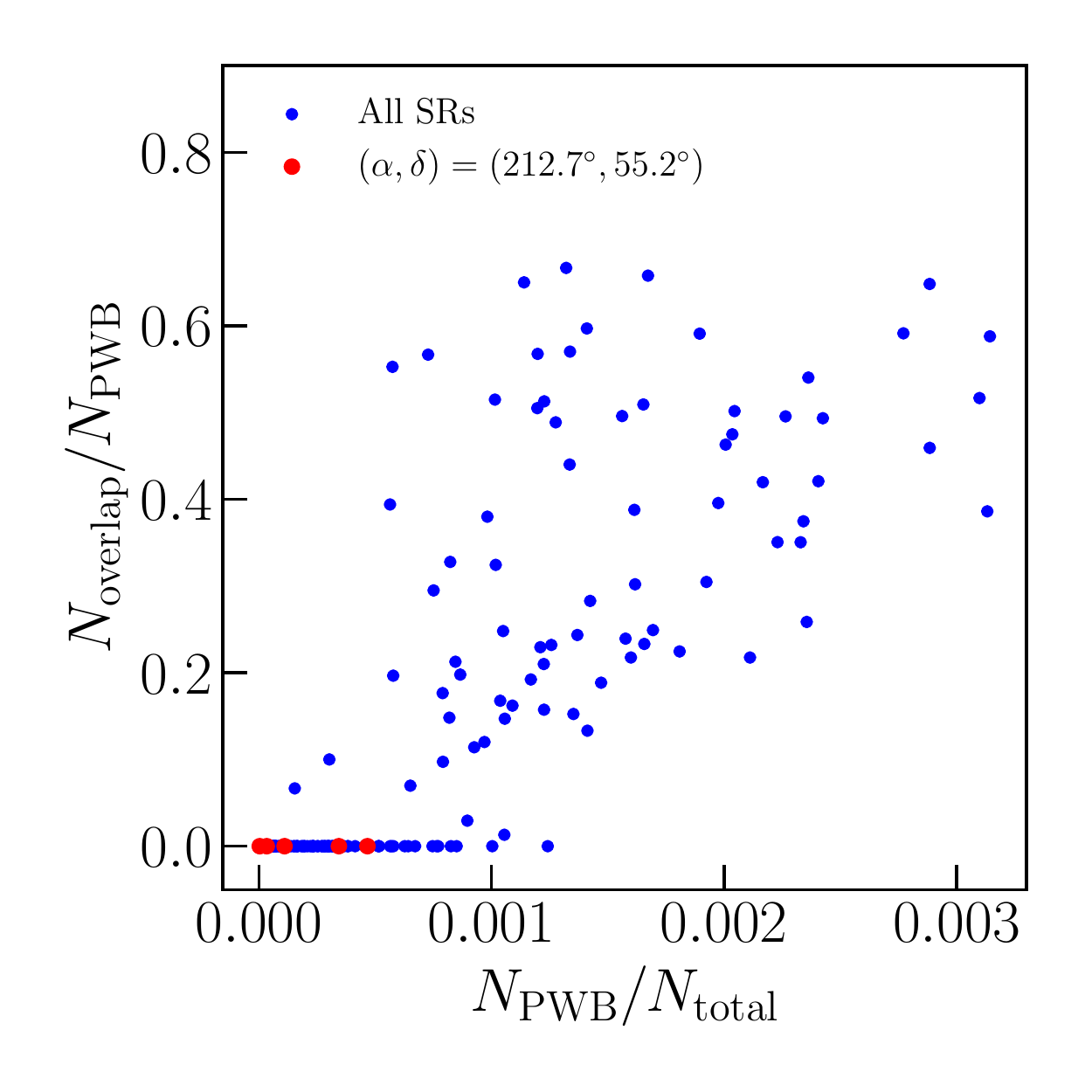}
\caption{For each SR, we plot the fraction of total stars $N_{\rm total}$ in an SR  which are identified as likely members of GD-1 by \citetalias{2018ApJ...863L..20P} ($N_{\rm PWB}$), compared to the fraction of $N_{\rm PWB}$ which are {\it also} identified by \textsc{Via Machinae} as likely members of GD-1 ($N_{\rm overlap}$). The SRs which lie in the patch centered on $(\alpha,\delta) = (212.7^\circ,55.2^\circ)$ are shown in red. This patch contains the majority of the right-hand side of GD-1 which is not identified in our analysis.} 
\label{fig:starfraction}
\end{centering}
\end{figure}

 Broadly speaking, we see that  \textsc{Via Machinae} has done an excellent job finding the GD-1 stars across the 21 patches of the sky considered in this work. Some notable features and caveats which deserve consideration are as follows:  
 \begin{itemize}
  \item  \Fig{fig:gd1_streamcoords} shows that \textsc{Via Machinae} has successfully reproduced some famous features of GD-1, including both gaps, the possible progenitor, and the ``spur" \citepalias{2018ApJ...863L..20P}.
  
  \item We see that \textsc{Via Machinae} confirms most of the additional 20$^\circ$ of GD-1 discovered in \citetalias{2018ApJ...863L..20P} (corresponding to $\alpha\lesssim -220^\circ$, or $\phi_1 \lesssim -60^\circ$). The left-most end of GD-1 ($\alpha\lesssim -235^\circ$, $\phi_1 \lesssim -80^\circ$) is missing from our stream candidate; this is because those patches were not included in our analysis as they were deemed too close to the disk ($|b|<30^\circ$). 
  
  \item On the right side of GD-1, we see that we are also missing stars compared to \citetalias{2018ApJ...863L..20P}. Closer inspection of this missing region reveals that this segment of the stream was captured by only a single patch, centered on $(\alpha,\delta) = (212.7^\circ,55.2^\circ)$, and the proper motion of GD-1 on this end of the stream is closer to $\mu_\lambda=0$, increasing the number of background stars in the relevant SRs. We will return to this point and elaborate on it further below. 
  
    \item The feature protruding from the stream at $\alpha\sim-215^\circ$ and $\delta\sim 40^\circ$ (see Fig.~\ref{fig:gd1_finalcluster}) is most likely an artifact of our line finding procedure.

  \item Of the 1,985 stars identified as likely members of GD-1 from \citetalias{2018ApJ...863L..20P}, 1,519 are in our fiducial (color and magnitude) region, and 738 (49\%) overlap with the membership of our stream candidate.

 \item The remaining 950 stars in our stream candidate were not tagged by \citetalias{2018ApJ...863L..20P}. Some of these may very well be additional members of GD-1. However, the proper motion and color-magnitude plots of Fig.~\ref{fig:gd1_finalcluster} make clear that our method also picks up a significant number of non-stream stars, see e.g.~the group of bright stars that are clearly not associated with the main GD-1 isochrone.
 \end{itemize}

 Regarding the stars missing from the right-most end ($\alpha \gtrsim -160^\circ$ or $\phi_1 \gtrsim -10^\circ$) of GD-1, it is notable that this side of the stream has values of $\mu_\lambda$ closest to zero (this is apparent in Fig.~\ref{fig:gd1_finalcluster} after taking into account that $\mu_\lambda\approx\mu_\delta$ for these patches). Hence, this segment of GD-1 falls primarily in SRs with an increased number of background stars compared to the rest of the stream. The fact that we do not recover this part of GD-1 strongly suggests that successful stream-finding with \textsc{Via Machinae} may require a minimum admixture of stream stars in the SR. In \Fig{fig:starfraction}, we use the stream candidates from \citetalias{2018ApJ...863L..20P} as a proxy for the GD-1 stars, and plot the fraction of \citetalias{2018ApJ...863L..20P}-tagged stars which are also identified as stream candidates by \textsc{Via Machinae} (in each SR) versus the fraction of stars in each SR which are tagged by \citetalias{2018ApJ...863L..20P}. As can be seen, the fraction of \citetalias{2018ApJ...863L..20P} stars also identified as stream members by \textsc{Via Machinae} is strongly correlated with the fraction of stream stars in the SR. In particular, the overlap fraction drops precipitously when the stream makes up $\lesssim 0.1\%$ of the total stars in the SR. All of the SRs through which the missing right-hand side of the GD-1 stream pass have a low fraction of stream stars. We believe this goes a long way toward explaining why \textsc{Via Machinae} missed these members of GD-1. Further work is needed (including a study of other streams beyond GD-1) to determine if this threshold is a more general requirement of ANODE and \textsc{Via Machinae} for stream detection. 
  
 Apart from the apparent required minimum $S/B$ detection thresholds for ANODE and \textsc{Via Machinae}, the fact remains that our stream candidate does not include all of the likely GD-1 stars tagged by \citetalias{2018ApJ...863L..20P} (completeness), and appears to include a substantial number of non-GD-1 stars (purity). However, this is not necessarily a drawback of the method. Rather, it reflects the emphasis placed by \textsc{Via Machinae} on {\it stream discovery} rather than {\it stream membership}. When designing our algorithm, our choices were motivated to identify stream candidates at a sufficiently high statistical significance to overcome the random background. Decisions such as the number of high-$R$ stars to include in each ROI and the line width in the Hough transform were made with this in mind, rather than maximizing accurate stream membership of the candidate. Loosening these criteria would likely recover more of the tagged stream stars than the 49\% identified here -- this would have to be weighed against increasing the number of false-positive stream candidates identified across the full sky.
 Thus, the resulting stream candidates should be taken as signs for discovery, rather than an accurate membership study of particular stars and whether they belong to a stream. After stream discovery, the candidate must be considered individually, loosening or eliminating some of the algorithmic choices that are part of \textsc{Via Machinae}. The density estimates from ANODE may continue to aid in this a posteriori analysis, but this is beyond the scope of the current paper.

\section{Conclusions }
\label{sec:conclusions}

In this work, we described a new machine learning-based algorithm called \textsc{Via Machinae} for stellar stream detection using {\it Gaia} DR2 data, and applied this  technique to identify the GD-1 stream. As a particularly distinct stream with readily available membership catalogs to use for detailed comparisons, GD-1 is an excellent testbed for our algorithm. 

The core of our technique is ANODE, a data-driven, unsupervised machine-learning algorithm that uses conditional probability density estimation to identify anomalous data points in a search region without having to explicitly model the background distribution. This approach is made possible by advances in deep learning that approximate the probability densities in an unsupervised way. We take as input for the ANODE training the angular position, proper motion, and photometry of the stars in {\it Gaia} DR2. No astrophysical knowledge is embedded into ANODE, other than in our choice to condition the probability estimation on one of the proper motion coordinates $\mu_\lambda$ -- which is also used to define the search regions. This allows us to identify potential anomalies while remaining agnostic to the Galactic potential, orbits, or stellar composition of the streams. 

The output of ANODE is a likelihood ratio $R$, with large $R$ values corresponding to stars whose phase space density in the search region is larger than expected based on interpolation from the control regions. To turn these anomalies into stream detection, \textsc{Via Machinae} engages in a number of additional steps. Some of these steps -- concentrating on old, metal-poor stars (identified with a cut on $b-r$, without requiring the stars to lie on an isochrone), further slicing the data into regions of interest based on the other proper motion coordinate $\mu_\phi^*$ -- are designed to improve signal-to-noise of stream detection, without sacrificing too much of the model-independence. Other steps -- automated line finding using the Hough transform, merging concordant best-fit lines in adjacent ROIs and patches of the sky -- are intended to build a stronger case for a stream detection (as opposed to some other anomalous structure or spurious false positive). The upshot is that \textsc{Via Machinae} produces a list of stream candidates that have been found in multiple overlapping search regions with high significance.

Using this method, we recover the GD-1 stream across the 21 patches in our full-sky scan that include it. Although GD-1 is an atypically dense, cold and narrow stream, it is still non-trivial that our method is able to recover it in an unsupervised and fully automated way. Moreover, we chose to focus on GD-1 in this paper as it provides a clear, step-by-step introduction to our algorithm. The application of \textsc{Via Machinae} to other known streams and to the full-sky dataset will be discussed in a forthcoming \citetalias{full_sky}. 

This initial application of unsupervised density estimators for stellar stream discovery, suggests other potentially interesting directions which recent advances in deep learning have made possible. Most obviously, our method can in principle be adapted to look for other interesting cold objects in the Milky Way, such as debris flow, tidal tails, and other stellar substructure \citep{1996ApJ...465..278J,1998ApJ...495..297J,2005ApJ...632..872R,2006ApJ...646..886F,2011MNRAS.416.2802F,2020ARA&A..58..205H}. Other methods of density estimation beside the MAF may also prove to be useful: we used the MAF because it is reasonably fast and easy to train and was demonstrated to perform well in the ANODE anomaly detection task in \citetalias{Nachman:2020lpy}. However, neural autoregressive flows \citep{huang2018neural}, neural spline flows \citep{durkan2019neural}, and mixture density networks \citep{bishop1994mixture} may possibly have improved performance in some or all contexts. 

Having data-driven measures of ``signal'' and ``background" densities may prove to be useful for problems beyond discovery. Sampling from these density estimators is possible, and might be a way to construct mock catalogues. The density estimates themselves might be useful for answer questions of stream membership. This could help in going beyond the \textsc{Via Machinae} discovery steps outlined in this paper, and further establish the validity and accuracy of the proposed stream candidates.

\section*{Acknowledgements}
\addcontentsline{toc}{section}{Acknowledgements}

We would like to thank A. Bonaca, D. Hogg, S. Pearson, A. Price-Whelan for helpful conversations; and Ting Li, Ben Nachman and Bryan Ostdiek for comments on the manuscript.
MB and DS are supported by the DOE under Award Number DOE-SC0010008.
LN is supported by the DOE under Award Number DESC0011632, the Sherman Fairchild fellowship, the University of California Presidential fellowship, and the fellowship of theoretical astrophysics at Carnegie Observatories.
LN is grateful for the generous support and hospitality of the Rutgers NHETC Visitor Program, where this work was initiated.

This research used resources of the National Energy Research Scientific Computing Center (NERSC), a U.S. Department of Energy Office of Science User Facility operated under Contract No. DE-AC02-05CH11231.

This work has made use of data from the European Space Agency (ESA) mission
{\it Gaia} (\url{https://www.cosmos.esa.int/gaia}), processed by the {\it Gaia}
Data Processing and Analysis Consortium (DPAC,
\url{https://www.cosmos.esa.int/web/gaia/dpac/consortium}). Funding for the DPAC
has been provided by national institutions, in particular the institutions
participating in the {\it Gaia} Multilateral Agreement.


\section*{Data Availability}

This paper made use of the publicly available \Gaia DR2 data. For the GD-1 stars identified through our analysis, please email the corresponding author.



\bibliographystyle{mnras}
\bibliography{streamfinding} 

\appendix

\section{Details of the MAF}

\subsection{Training and model selection}
\label{app:training}

For each SR -- defined by a patch of the sky and a slice in $\mu_\lambda$ -- we train two separate MAFs, one on the stars $\mu_\lambda\in [\mu_\lambda^{\rm min},\mu_\lambda^{\rm max}]$ in the SR and one on the stars in its complement (the CR), $\mu_\lambda\notin [\mu_\lambda^{\rm min},\mu_\lambda^{\rm max}]$. Before training, the data is standardized by shifting the mean in each feature to zero and normalizing the standard deviation to unity.

We opt not to divide the data up into training and validation sets, as doing so would dilute the significance of any stream detection. Based on direct inspection, we do not find any evidence for overfitting. For the density estimation, overfitting would typically correspond to $p(x)$ degenerating into a set of delta functions centered on each point in the training data. This is generally not a concern for the MAF, and in fact it generally has the opposite problem (not being able to fit extremely sharp distributions). Also, the lower bound on dataset size (SRs must have at least 20,000 stars, otherwise they are rejected) should be sufficient to mitigate overfitting for the dimensionality of the feature space.

For each MAF, we train for 150 epochs using the Adam optimizer \cite{adam}. The learning rate is a hyperparameter that will be included in the scan to be described in \Sec{app:hyperparam}. This number of epochs seemed to be sufficient for convergence, and training for significantly longer is computationally prohibitive. To smooth out fluctuations in the MAF from epoch to epoch arising from stochastic gradient descent, we  calculate a running average for each star's probability density over the output of 20 consecutive training epochs. 

To select the best model for each MAF, we employ the following approach. On general grounds, we expect the $\log R$ distribution to be roughly symmetric around 0 in the absence of any signal; any deviation from $R=1$ is due to random fluctuations in the MAFs estimating the numerator or the denominator of the likelihood ratio. The better the performance of the density estimation, the more sharply peaked the $R$ distribution should be around $R=1$. Furthermore, we expect astrophysical signals (such as streams) to typically correspond to {\it over}densities, not underdensities. Putting all this together, we select the ``best" epoch by considering the $\log R$ distribution for $\log R<0$, reflecting this across 0, and choosing the epoch with the smallest standard deviation in this symmetrized distribution. Strictly speaking, we only perform this for the MAF trained on the CR; for the MAF trained on the SR, we take the last 20 epochs, as we found this led to the best performance on tests with the labelled GD-1 stars.

\subsection{Hyperparameter Optimization}
 \label{app:hyperparam}
 
Here we describe the hyperparameter optimization for the MAF neural network used for density estimation in this work. For the MAF architecture, these hyperparameters include the number of blocks in the neural network that make up the affine transformations, the number of hidden layers in the network, and the number of nodes in each hidden layer. Then there are the usual hyperparameters involved in training (mini-batch size, learning rate, etc.).
The optimal values for these hyperparameters are not derivable from first principles; instead, we must perform a scan over the hyperparameters and select the configuration that maximizes the performance of the neural network. To measure performance, we will use the GD-1 labelled stars from \citetalias{2018ApJ...863L..20P} and quantify the signal/background discrimination power of the ANODE method as in Sec.~\ref{sec:anode}.

To optimize the hyperparameters, we used a $15^\circ$ patch of the sky centered on $(\alpha,\delta) = (140^\circ,30^\circ)$, which contains a segment of the GD-1 stream (this patch was hand-selected, and is not one of the 200 centers described in Sec.~\ref{sec:inputs}). The patch contains $1.2\times 10^6$ stars, of which 574 were tagged as stream stars by \citetalias{2018ApJ...863L..20P}. The inner $10^\circ$ fiducial region has $4.3\times 10^5$ stars, 374 of which are stream-tagged.

Using these tagged stars, we hand-pick an SR defined by $\mu_\alpha^* \in [-8.75,-15]~{\rm mas/yr}$, which contains all the stream stars and $1.7\times 10^5$ total stars ($6.4\times 10^4$ in the fiducial region).

\begin{figure}
\includegraphics[width=0.9\columnwidth]{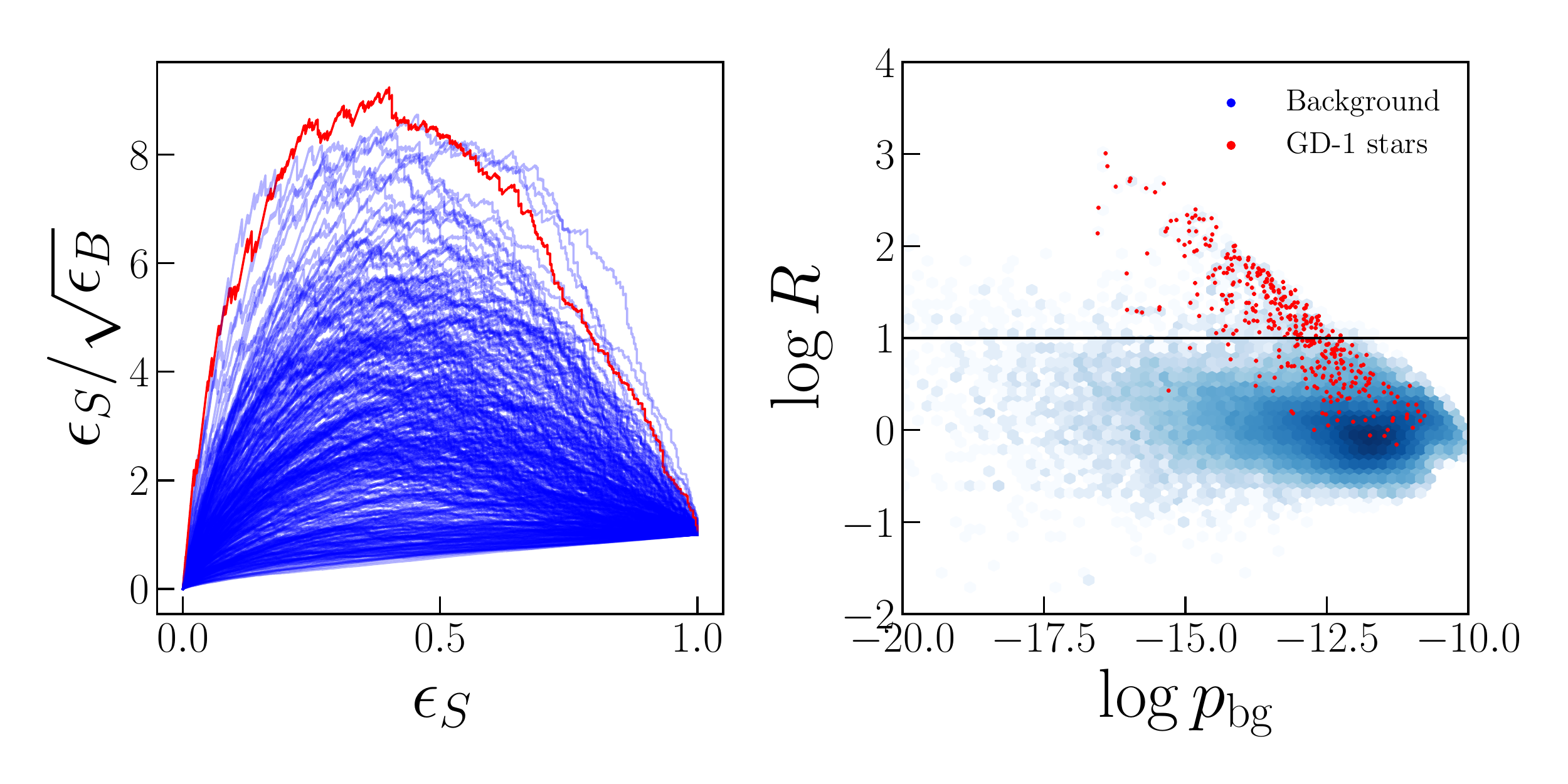}
\caption{Left: SIC curve of signal efficiency $\epsilon_S$ to $\epsilon_S/\sqrt{\epsilon_B}$ (for a background efficiency $\epsilon_B$) as a cut is placed on $\log R$, for all hyperparameters tested on the GD-1 example dataset. Right: Density plot of $\log p_{\rm bg}$ versus $\log R$ for stars in the signal region of the GD-1 dataset used for hyperparameter optimization, trained using the neural network parameters that maximize the true-positive over root false-positive rate.
\label{fig:GD1example_R}}
\end{figure}

We varied the hyperparameters over batch size, number of epochs, learning rates, number of blocks, and number of hidden layers with:
\begin{eqnarray}
{\rm batch~size} & = & [256,512,1024] \nonumber \\
{\rm num.~blocks} & = & [16,18,20,25] \nonumber \\
{\rm num.~hidden} & = & [16,32,64] \\
{\rm num.~epochs} & = & [125,150,175] \nonumber \\
{\rm learning~rate} & = & [5\times 10^{-5},7\times 10^{-5},2\times 10^{-5},4\times 10^{-5}]. \nonumber
\end{eqnarray}
 We train the MAF for each combination of these five parameters. For each hyperparameter set in the scan, we calculate a $\log R$ value for each star in the fiducial (inner $10^\circ$) region. After training, we use the last epoch to construct the significance improvement characteristic (SIC) curve by varying a cut on $\log R$. The SIC curves for each hyperparameter configuration in the scan are plotted in the left panel of \Fig{fig:GD1example_R}, with the optimal choice that maximizes $\epsilon_S/\sqrt{\epsilon_B}$ highlighted in red. On the right panel of Figure~\ref{fig:GD1example_R}, we show the distribution of $\log p_{\rm bg}$ versus $\log R$ for stars in the SR for the optimal set of hyperparameters. The optimal hyperparameters -- 150 epochs, a batch size of 512, 20 blocks, 64 hidden blocks, and a learning rate of $7\times 10^{-5}$ -- are then  used for all MAF trainings in this work.

\section{Globular cluster detection}
\label{app:GC}

Here we describe the simple algorithm we use to remove SRs  that contain a suspected globular cluster. The presence of such overdensities in an SR is enough to distort the density estimation; the MAF cannot fit the delta function that is a globular cluster while simultaneously accurately describing the rest of the patch. 

Based upon inspecting many patches pre- and post-ANODE, we find that the GCs that spoil the MAF are usually visible as a single bright pixel in a simple two-dimensional (2D) density plot of the stars' positions in an SR.   
Given that there are thousands of SRs to sift through, we make a 2D histogram of the latitude and longitude of all the stars in an SR (recall, the patch size is 15$^\circ$). With some tuning, we find that a good resolution is $120\times 120$ bins across the $15^\circ\times 15^\circ$ region. We then compute the mean number of counts $\bar N$, the max number of counts $N_{\rm max}$, and the standard deviation of the number of counts (as measured by the inter-quartile range) $\sigma$. We declare the SR to contain a likely GC if 
\beq
{{N_{\rm{max}}-\bar N}\over \sigma} > 4 \qquad {\rm and}\qquad N_{\rm{max}}>25.
\eeq
In other words, the bin with the maximum number of stars had to have at least 25 stars, and had to be at least ``4$\sigma$" significant over the background stellar distribution. 

Using these simple criteria, we find 1,381 (out of 6,117) SRs in the full-sky dataset contain a GC candidate. We have visually inspected all of the SRs containing GC candidates and confirmed that the selections appear to be reasonable.

\section{Comments on stream 2 and disk stars}
\label{sec:stream2}

\begin{figure*}[b!]
\begin{centering}
\includegraphics[width=1.8\columnwidth]{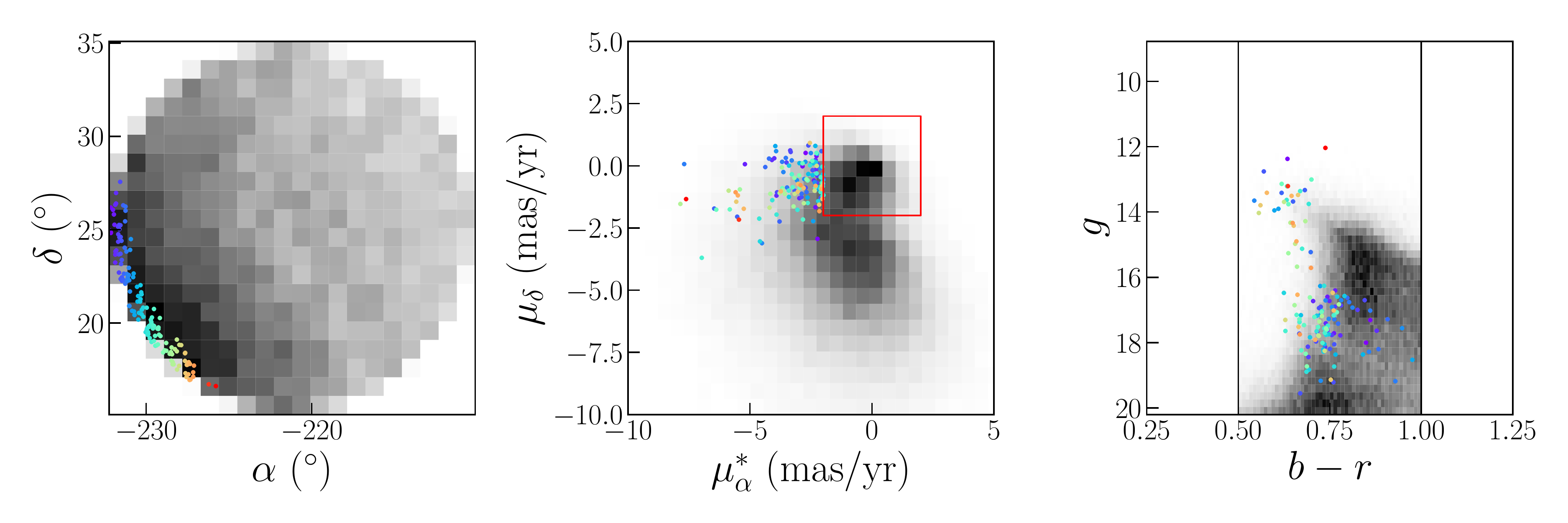}
\caption{Scatter plots of the angular positions, proper motions, and color/magnitudes of the stars in the second, less prominent stream candidate identified by \textsc{Via Machinae}, overlaid on 2d histograms of all the stars in the circular patch that contains this stream candidate (darker pixels indicate higher density of stars). As in Fig.~\ref{fig:gd1_finalcluster}, the \textsc{Via Machinae} stars are color-coded by position in $\alpha$, to facilitate cross referencing between the three individual scatter plots.}
\label{fig:2ndstream}
\end{centering}
\end{figure*}

Here we elaborate further on the second, less prominent stream candidate tagged by the full \textsc{Via Machinae} algorithm in the 21 patches containing GD-1. As described in Sec.~\ref{sec:gd1}, this stream candidate is contained completely in a single patch (centered on $(\alpha,\delta)=(138.8^\circ,\,25.1^\circ)$), and all of the high-$R$ stars follow tightly the edge of the circular patch, aligned and on the same side as the Galactic disk. We illustrate this further in Fig.~\ref{fig:2ndstream}, which show the stars of the stream candidates in angular position, proper motion and color/magnitude space, overlaid on top of density plots of all the stars in the patch containing the stream candidate. This shows clearly how the stream candidate is aligned with the density gradient in the patch (which in turn is aligned with the Galactic disk, which one can check by transforming to Galactic coordinates $(\ell,b)$). We also see that the stream candidate is clustered in proper motion space close to $\mu_\alpha^*,\,\,\mu_\delta\sim 0$, which as we have noted in Sec.~\ref{sec:roi} is a significant source of false positives for the ANODE method. Finally, we note that (unlike for GD-1 and other known streams), there is no noticeable correlation between the position along the stream and the proper motion. Taken together, we view this as strong evidence that this second stream candidate is likely to be a false positive.

More generally, we observe a strong gradient in stellar density towards the Galactic disk in many patches and SRs. There is also likely a strong correlation between disk stars and proper motion within a patch.\footnote{Understanding this correlation requires modeling stellar orbits in the Milky Way, and a detailed understanding of projection and line-of-sight effects. This is beyond the scope of the present work.} 
Therefore, it is potentially concerning that \textsc{Via Machinae} could systematically misidentify disk stars as stream stars. 

A conservative approach to avoid this misidentification is to reject all ROIs where the line-finder returned best fit parameters that are at the edge of the patch closest to the Galactic disk and parallel to it. Specifically, we propose to cut out all ROIs whose best-fit line radius has $|\rho|>9.5^\circ$, slope less than 0.2 radians in Galactic $\ell$, $b$ coordinates (that is, aligned with the disk), and are localized on the side of the patch nearest to the disk. This requirement removes only 91 ROIs (out of $\approx 17,000$) from our sample. Such cut would eliminate the second stream that we find in \Sec{sec:gd1}, but it would not affect the GD-1 stream candidate at all.



\bsp	
\label{lastpage}
\end{document}